\documentclass[aps, prd,eqsecnum,superscriptaddress,twocolumn,
nofootinbib,floatfix,preprintnumbers]{revtex4-1}

\usepackage{graphicx}
\usepackage{amssymb}
\usepackage{amsmath}
\usepackage{rotating}
\usepackage{color}
\usepackage{soul}
\usepackage{aeitensor}
\usepackage{bm}
\usepackage{rcs}
\usepackage{pbox}
\usepackage{tabularx}
\usepackage{mathtools}
\usepackage[mathscr]{euscript}
\usepackage{multirow}
\usepackage[caption=false]{subfig}

\newcolumntype{P}[1]{>{\centering\arraybackslash}m{#1}}

\makeatletter
\newcommand\footnoteref[1]{\protected@xdef\@thefnmark{\ref{#1}}\@footnotemark}
\makeatother

\renewcommand{\today}{\number\year\space\ifcase\month\or
  January\or February\or March\or April\or May\or June\or
  July\or August\or September\or October\or November\or December\fi
  \space\number\day}

\newcommand{\Tobs}{T_{\text{obs}}}
\newcommand{\Tcoh}{T_{\text{coh}}}

\newcommand{\Npairs}{N_\text{pairs}}
\newcommand{\Nsft}{N_\text{SFT}}
\newcommand{\Ncoh}{N_\text{coh}}
\newcommand{\Nbin}{N_\text{bin}}
\newcommand{\Ndet}{N_\text{det}}
\newcommand{\hr}{h_\text{rms}}
\newcommand{\baseline}{\Delta T_{\text{SFT}}}
\newcommand{\scale}{C_\chi}

\newcommand{\ncp}{\frac{\hr^2\Tobs (\A^2_+F^2_{+}+\A^2_\times F^2_{\times})}{S_n}}

\newcommand{\y}{\mathcal{Y}}

\newcommand{\G}{\tilde{\mathcal{G}}}
\newcommand{\x}{\tilde{x}}
\newcommand{\A}{\mathcal{A}}

\newcommand{\unit}[1]{~\mathrm{#1}}
\newcommand{\normcdf}{\mathscr{F}_\mathcal{N}}
\newcommand{\chicdf}{\mathscr{F}_\chi}
\newcommand{\ncxcdf}{\mathscr{F}_{\textsc{nc}\chi}}

\newcommand{\mflim}{118\,Mpc}

\DeclareMathOperator{\Cov}{Cov}

\begin{document}
\newcommand*{\TTU}{Department of Physics, Texas Tech University, Lubbock, TX 79409-1051 (USA)}\affiliation{\TTU}
\newcommand*{\GWU}{Department of Physics, George Washington University, Washington, DC 20052 (USA)}\affiliation{\GWU}
\newcommand*{\pennp}{Department of Physics, The Pennsylvania State University, University Park, Pennsylvania 16802-6300 (USA)}\affiliation{\pennp}
\title{Cross-correlation method for intermediate-duration gravitational wave searches associated with gamma-ray bursts}

\author{Robert Coyne}\email{rob.coyne@ttu.edu}\affiliation{\TTU}\affiliation{\GWU}
\author{Alessandra Corsi}\affiliation{\TTU}
\author{Benjamin J. Owen}\affiliation{\TTU}\affiliation{\pennp}
\date[\relax]{compiled \today }
\begin{abstract}
Several models of gamma-ray burst progenitors suggest that the gamma-ray event
may be followed by gravitational wave signals of $10^3$--$10^4$ seconds
duration (possibly accompanying the so-called X-ray afterglow ``plateaus'').
We term these signals ``intermediate-duration'' because they are shorter than
continuous wave signals but longer than signals traditionally considered as
gravitational wave bursts, and are difficult to detect with most burst and
continuous wave methods. The cross-correlation technique proposed by
[S. Dhurandhar et al., Phys. Rev. D \textbf{77}, 082001 (2008)], which so far
has been used only on continuous wave signals, in principle unifies both burst
and continuous wave (as well as matched filtering and stochastic background)
methods, reducing them to different choices of which data to correlate on which
time scales. Here we perform the first tuning of this cross-correlation
technique to intermediate-duration signals. We derive theoretical estimates of
sensitivity in Gaussian noise in different limits of the cross-correlation
formalism, and compare them to the performance of a prototype search code on
simulated Gaussian-noise data. We estimate that the code is likely able to
detect \emph{some} classes of intermediate-duration signals (such as the ones described in
[A. Corsi \& P. M\'esz\'aros, Astrophys. J., \textbf{702}, 1171 (2009)]) from sources located 
at astrophysically-relevant distances of several tens of Mpc.
\end{abstract}
\maketitle
\section{Introduction}
\label{sec:intro}
Over the last decade, the LIGO and Virgo gravitational wave (GW) detectors
have carried out triggered (or targeted) GW searches in coincidence with
Gamma-Ray Bursts (GRBs) and other electromagnetic transients
\cite{Abbott2005,Abbott2007b,Acernese2007,Abbott2008a,Abbott2008b,Acernese2008,
Abbott2009c,Abbott2010a,Abbott2010b,Abadie2011a,Abadie2011b,Abbott2012,
Aasi2013c,Aasi:2014ent,Aasi:2014iia} as well as persistent electromagnetic
sources \cite{Abbott:2003yq,Abbott:2004ig,Abbott2007a,Abbott2007c,Abbott2008c,
Abadie2010,Abadie2011c,Abadie2011d,Aasi2013b,Aasi:2013sia,2015PhRvD..91b2004A,
Aasi:2014qak,Aasi:2014ksa}. These searches have traditionally been optimized to
detect well-modeled ``chirp'' signals from neutron star (NS)-NS and/or
black-hole (BH)-NS binary inspirals, unmodeled short ($\lesssim 1-10$\,s)
duration bursts of GWs in association with electromagnetic transients, and
persistent (continuous) GWs from nearby rotating NSs. Searches based on methods
for a stochastic background have also been adapted to continuous wave
targets~\cite{Abbott2007d, Abadie2011d}. 

Methods targeting an as of yet largely unexplored class of ``intermediate
duration'' GW signals have also been developed \cite{Thrane2011,
Thrane2014,Thrane:2015wla} and two so far have led to a search on real data
\cite{Aasi2013c, Abbott2015}.\footnote{Those works use ``long'' to refer to signals of
$\mathcal{O}(10^2)\,\rm{s}$ duration, because these durations are long
compared to the $\mathcal{O}(\lesssim1)\,\rm{s}$ duration signals
traditionally targeted in burst data analyses. The term ``very long duration''
signals has also been adopted to refer to GWs lasting from hours to weeks,
e.g. \cite{Thrane2015}. Here, we use ``intermediate'' to put the discussion in
the broader context, which includes the substantially longer continuous wave
signals.} Intermediate duration GWs are of special interest in several
astrophysical scenarios (e.g.,
\cite{vanPutten2001,DallOsso2007,Piro2007,Corsi2009,vanPutten2008,Ott2009,Piro2011,Piro2012,Aasi2013c,
Doneva:2015jaa}), and their detectability over a large parameter space remains mostly unexplored compared to the more traditional inspiral, burst, or continuous wave signals.

In this work, we focus on the possibility of detecting $10^3-10^4$\,s duration GWs in coincidence with GRBs. Our study is motivated by the need for a data analysis technique that is optimized to probe some of the  long-lived progenitor scenarios for (long and short) GRBs, such as the so-called ``magnetar model''.  The magnetized NS (magnetar) scenario has been invoked to explain X-ray ``plateaus'' ($10^2-10^4$\,s-long periods of relatively constant emission) observed in $\gtrsim 50\%$ of long, and in several short, GRB afterglows \cite{Nousek2006,Zhang2006,Liang2007,Starling2008,Bernardini2012,Gompertz2013,Rowlinson2013,Yi2014}. Gravitational collapse leading to the formation of a NS, in turn, has long been considered an observable source of GWs. In a rotating, newly born NS, non-axisymmetric instabilities such as the secular Chandrasekhar-Friedman-Schutz \citep[CFS,][]{Chandra70,Friedman78} instabilities, can yield GW emission with high efficiency \cite{LaiShapiro1995}. If the newly born GRB-magnetar emits GWs over the plateau timescale ($\sim$$10^3$\,s), GW detectors such as the
advanced LIGO (aLIGO) and Virgo detectors may be able to directly probe the source of the observed prolonged energy injection, and clarify one of the key open questions on the nature of GRB central engines  \cite{Zhang2001,Corsi2009}.

Detecting intermediate-duration GW signals, such as the ones possibly associated with GRB plateaus, requires search techniques that can bridge the gap (both in terms of science reach and signal detection strategies) between traditional inspiral/burst searches, and continuous wave or stochastic ones.  Traditional short duration inspiral and long duration continuous wave searches make use of highly sensitive coherent (and computationally limited semi-coherent) techniques that leverage accurate knowledge of the expected GW waveform (as a function of a set of physical parameters). Traditional burst and stochastic searches, on the other hand, assume little a priori knowledge of the signal and depend respectively on excess signal power (above the background noise) and cross-correlation of power between interferometers for detection.

Here we address the problem of searching for intermediate-duration, large frequency bandwidth signals by adapting the cross-correlation method of \citep{Dhurandhar2008}. While originally developed in the context of continuous waves, the method by \citep{Dhurandhar2008} encompasses all of the aforementioned traditional search techniques when various parameters are taken to the appropriate limits, and it shows how to make best use of the information available about each type of signal. (A Bayesian framework of similarly broad relevance was developed later in \citet{Cornish:2013nma}, but here like \citet{Dhurandhar2008} we present an essentially frequentist analysis.) We correct some small errors in the original formalism of  \citep{Dhurandhar2008}, and apply it for the first time to intermediate-duration signals by developing a code, the performance of which we tested on simulated data. We restrict ourselves to intermediate-duration signals with large frequency bandwidth (such as the ones described in \cite{Corsi2009}), since intermediate-duration narrow band signals have different astrophysical origins and are treated with adaptations of continuous wave searches (see e.g. \citep{Prix:2011qv}).

Our paper is organized as follows. In Sec. \ref{sec:motivation} we motivate the application of Dhurandhar et al.'s \citep{Dhurandhar2008}  cross-correlation technique to intermediate-duration GWs. In Sec. \ref{sec:notation} we describe our notation and assumptions. In Sec. \ref{sec:xcorr} we briefly re-derive the general statistical behavior of the cross-correlation method, discuss explicitly its limits and intermediate regimes, and show how several assumptions made in \citep{Dhurandhar2008} need to be modified for the search of non well-modeled GW transients evolving on $10^3-10^4$\,s timescales. In Sec. \ref{sec:plateau} we apply the
cross-correlation technique to the model of secularly unstable GRB-magnetars described in \citep{Corsi2009}, thus providing an example of applicability to astrophysically motivated waveforms of intermediate-duration. Finally, in Section \ref{sec:discussion}, we compare our results
with other data analysis techniques that have been proposed to search for intermediate-duration GW signals, and give our conclusions.
\section{Motivation for a cross-correlation search}
\label{sec:motivation}

GWs signals are typically predicted to have strengths so close to the level of noise in the detectors that it is necessary to filter the interferometer data streams to detect the real GW events amongst spurious noise events. When the functional form of the predicted GW signal is very well known (as a function of a set of physical parameters), matched filtering with template waveforms is the optimal strategy (e.g., \cite{Schutz1991,Jar1998}). Matched filtering involves computing the cross-correlation between the interferometer output and a template waveform, weighted inversely by the noise spectrum of the detector. The signal-to-noise ratio (SNR) is defined as the cross-correlation of the template with a particular stretch of data divided by the root-mean-squared (rms) value of the cross-correlation of the template with pure detector noise.

Usually, a family of templates spanning the possible range of parameter values (a so-called template bank) is used in real data analyses.  A template bank adds to the search statistics a trial factor, which has to be taken into account when estimating the detection sensitivity. A template bank also involves more computational cost since each template must be cross-correlated with the data. While the parameters describing the search templates typically vary continuously throughout a finite range of values, a realistic template bank is composed of templates, the parameter values of which vary in discrete steps within the allowed range. The ``mismatch'' between the signal and nearest of the discrete templates causes some reduction in the expected matched filter SNR. Thus, the number of templates to be used in a search is a compromise between the maximum computational cost one can sustain, and the maximum mismatch that one is willing to tolerate (e.g., \cite{Owen1996,Brady1998,Owen1999,Wette2008}).

When the maximum sustainable computational cost implies a mismatch such that the loss in SNR reduces the sensitivity of the search to a very limited portion of the parameter space, modifications to the matched filtering strategy toward sub-optimal techniques are mandatory. In addition, in many cases, the GW signal waveform is not known well enough for matched filtering.  Indeed, even if a very finely spaced discrete template bank is used, a search may fail to detected a signal if the templates do not represent with sufficient accuracy the relevant physics. In other words, a realistic search is affected not only by the mismatch but also by the so-called ``fitting factor'' \citep{Apostolatos1995,Lee2008,Lee2010,Damour2011}, the fractional loss in SNR caused by the fact that even the best template in a family is only a ``fit'' to a hypothetical exact gravitational waveform. In the context of GWs from compact binaries, where numerical relativity can be used to quantify the fitting factor of phenomenological waveforms used to construct template banks for matched filter searches (e.g., \citep{Hinder2013}), it has been estimated that fitting factors $< 3\%$ are needed to achieve detection efficiencies $>90\%$ (see e.g. \citep{Apostolatos1995,Samson2014}). Indeed, matched filtering is by construction highly likely to miss a signal even for moderately bad fitting factors. On the other hand, sub-optimal (less sensitive) detection techniques are more robust against the intrinsic uncertainties in the underlying physics \citep{Brady2000,Allen2005,Cutler2005}.

In the case of secular bar-mode GW signals from GRB afterglow plateaus, given the uncertainties related to the physics of GRB central engines, the derived gravitational waveforms are to be considered as simplified phenomenological models. Thus, a more robust (when compared to matched filtering) search is necessary. A very robust approach against signal uncertainties consists of using the cross-correlation between the output of \textit{different, non-colocated} detectors. This approach (which, differently from matched filtering, requires no a-priori knowledge of the signal waveform and its properties) is typically used for stochastic GW background searches (e.g., \cite{Abbott2007d,Abbott2009b,Abadie2014}). The cross-correlation between different, non-colocated detectors, only relies on the fact that, in the presence of a GW signal, the output from distinct detectors (at the same times, after correcting for the light-travel time between detectors) should be correlated, while pure noise would remain uncorrelated. Of course, this technique also implies a poor resolution in the parameter space, and more expensive follow-ups to verify possible detections \citep{Dhurandhar2008}.

It is important to note that the cross-correlation is at the basis of two opposite search strategies: the (highly sensitive) matched filtering (cross-correlation of the data with a template), and the (very robust) ``stochastic search'' (cross-correlation of different detectors' output). Indeed, by noticing this fundamental fact, Dhurandhar et al. 2008 \citep{Dhurandhar2008} have provided an elegant formulation of the cross-correlation statistic for periodic GW searches such that, depending on the maximum duration over which one believes phase coherence is preserved by the signal, the statistic can be tuned to go from a ``stochastic-type'' search using data from distinct detectors, to the semi-coherent time-frequency methods with increasing coherent time baselines (e.g., \cite{Brady1998}), and all the way to a fully coherent search (nearly recovering the matched filtering statistic).

Dhurandar et al.'s formulation of the cross-correlation statistic \citep{Dhurandhar2008} leads to a unified framework that can be used to make informed trade-offs between computational cost, sensitivity, and robustness against signal uncertainties.  Studies based on the cross-correlation statistic as formulated by \cite{Dhurandhar2008} have focused on continuous GW emission from Supernova 1987a and Scorpius X-1 \citep{Chung2011,Whelan2015}, and a number of refinements to the cross-correlation method have also been published in recent years, particularly for the treatment of spectral leakage \citep{Sundaresan2012, Whelan2015}. In what follows, we present a strategy tuned for the detection of \textit{intermediate-duration} ($\lesssim 10^{4}$\,s) quasi-periodic GW signals, and discuss its application to the case of secularly unstable GRB magnetars (Section \ref{sec:plateau}).

\section{Notation and Assumptions}
\label{sec:notation}
\subsection{The Short-time Fourier Transform}
The Short-time Fourier Transform (SFT) is a useful tool when examining a signal in which frequency content is evolving with time. The time-domain output of LIGO/Virgo detectors, $x(t)$, can be represented as the linear combination of a GW signal $h(t)$, and background noise $n(t)$:
\begin{equation}
x(t)=h(t)+n(t).
\label{eq:detector-output}
\end{equation}
The SFT of the detector output is constructed by dividing the time-series $x(t)$ into $\Nsft$ segments of duration $\baseline$ (generally speaking, these segments may or may not overlap), and by taking the Discrete Fourier Transform (DFT) of each of these segments:
\begin{equation}
\tilde{x}_I[f_k] = \frac{1}{f_s}\sum\limits_{l=0}^{\Nbin-1} x[t_l]\mathrm{e}^{-2\pi i f_k (t_l-T_I+\baseline/2)},
\label{eq:dft}
\end{equation}
where $f_s$ is the sampling frequency (typically $f_s=16,384$\,Hz for the LIGO detectors); $\Nbin=\baseline\times f_s$ is the number of frequency bins of each SFT; and $f_k$ is the frequency corresponding to the $k$-th frequency bin:
\begin{eqnarray}
f_k=\frac{k}{\baseline}~~~~{\rm for}~~~ k=0,...,\Nbin/2-1,~~~\\
f_k=\frac{(k-\Nbin)}{\baseline}~~~~{\rm for}~~~ k=\Nbin/2,...,\Nbin-1.~~
\end{eqnarray}
Note that $t_l$ in Eq. (\ref{eq:dft}) corresponds to the $l$-th time sample i.e., $t_l=T_I-\baseline/2 + l/f_s$. For each $I=0, 1, ... \Tobs/\baseline$ (where $\Tobs$ is the total duration of the signal) and $l=0,1,...,\Nbin$, $t_l$ spans the time interval $T_I-\baseline/2\le t_l \le T_I+\baseline$. Note also that we distinguish between continuous time series $x(...)$ and their associated discretely-sampled time series $x[...]$ by using square brackets.

To reduce spectral leakage, a windowing function $w[t_l]$ is often applied to the DFT \cite{Percival1993}:
\begin{equation}
\tilde{x}_I[f_k] = \sum\limits_{l=0}^{\Nbin-1} w[t_l]x[t_l]\mathrm{e}^{-2\pi i f_k (t_l-T_I+\baseline/2)}.
\label{eq:windowed-dft}
\end{equation}
For simplicity, and following \cite{Dhurandhar2008}, hereafter we neglect the window function (but discuss some of the related issues in Section \ref{sec:specleak}).

\subsection{Detector noise and its PSD}
\label{sec:sft}
In this Section we consider the detector output in the absence of a signal. In the continuum limit of Eq. \eqref{eq:detector-output}, the frequency ($f$) content of the detector noise can be described by its Fourier transform:
\begin{equation}
\tilde{n}(f)=\int\limits_{-\infty}^\infty dt~n(t)\mathrm{e}^{-2\pi i ft}.
\label{eq:noise-transform}
\end{equation}
The single-sided ($f\gtrsim0$) Power Spectral Density (PSD) of the noise, $S_n(f)$, is defined as:
\begin{equation}
S_n(f):=2 \int\limits_{-\infty}^\infty d\tau~\langle n(t)n(t+\tau)\rangle \mathrm{e}^{-2\pi i f\tau},
\label{eq:PSD}
\end{equation}
where $\langle n(t)n(t+\tau) \rangle$ is the autocorrelation function of the noise, and the expectation value $\langle \cdot \rangle$ represents an average over an ensemble of noise realizations. The noise autocorrelation function thus forms a Fourier transform pair with its PSD. Note that hereafter we assume the noise is stationary and Gaussian (with zero mean), thus its autocorrelation function is independent of $t$.

From Eq. \eqref{eq:noise-transform}, it follows that (see also \cite{Creighton2011}):
\begin{equation}
\langle \tilde{n}^*(f')\tilde{n}(f)\rangle=\left\langle\int\limits_{-\infty}^\infty dt'~n^*(t')\mathrm{e}^{2\pi i f't'}\int\limits_{-\infty}^\infty dt~n(t)\mathrm{e}^{-2\pi i ft}\right\rangle .
\end{equation}
This product of independent integrals can be recast as:
\begin{equation}
\langle \tilde{n}^*(f')\tilde{n}(f)\rangle=\left\langle\int\limits_{-\infty}^\infty dt'\int\limits_{-\infty}^\infty dt~n^*(t')n(t)\mathrm{e}^{2\pi i f't'}\mathrm{e}^{-2\pi i ft}\right\rangle.
\end{equation}
Noting that real detector output implies $n^*(t)=n(t)$, and given the linearity and limited multiplicativity\footnote{The expectation value $\langle XY\rangle$ of random variables $X,~Y$ is multiplicative if $\Cov(X,Y)=0$. That is, only if $X$ and $Y$ are statistically independent.}of the expectation value, we have:
\begin{equation}
\langle \tilde{n}^*(f')\tilde{n}(f)\rangle=\int\limits_{-\infty}^\infty dt'\int\limits_{-\infty}^\infty dt~\left\langle n(t')n(t)\right\rangle\mathrm{e}^{2\pi i f't'}\mathrm{e}^{-2\pi i ft}.
\end{equation}
Setting $t=t'+\tau$, yields:
\begin{multline}
\langle \tilde{n}^*(f')\tilde{n}(f)\rangle=\\
\int\limits_{-\infty}^\infty dt'~\mathrm{e}^{-2\pi i (f-f')t'}\int\limits_{-\infty}^\infty d\tau~\langle n(t')n(t'+\tau)\rangle\mathrm{e}^{-2\pi i f\tau}.
\end{multline}
Then, using Eq. (\ref{eq:PSD}), we replace the integral over $d\tau$ with the PSD,
\begin{equation}
\langle \tilde{n}^*(f')\tilde{n}(f)\rangle=\frac{S_n(f)}{2}\int\limits_{-\infty}^\infty dt'~\mathrm{e}^{-2\pi i (f-f')t'}.
\end{equation}
The remaining integral over $dt'$ is simply a delta function,
\begin{equation}
\langle \tilde{n}^*(f')\tilde{n}(f)\rangle=\frac{1}{2}\delta(f-f')S_n(f),
\end{equation}
and using the finite time approximation of the delta function:
\begin{equation}
\delta_{\baseline}(f)=\frac{\sin(\pi f \baseline)}{\pi f},
\label{eq:finite-delta}
\end{equation}
which reduces to $\baseline$ in the limit of $f\rightarrow0$, we can relate the variance of the Fourier transformed detector output to the PSD:
\begin{equation}
\langle |\tilde{n}_I[f_k]|^2 \rangle \approx \frac{\baseline}{2}S_n[f_k].
\label{eq:noisepsd}
\end{equation}

\subsection{Short-duration Fourier Transform of the signal}
\label{signalsft}
We make the hypothesis that the GW signal $h(t)$ is quasi-periodic (by taking a sufficiently small time interval the signal in such an interval can be considered monochromatic), and assume that its time-frequency evolution is described with sufficient physical accuracy, for a time interval of $\Tcoh$, via some known function of a given set of parameters (although this function may not have a closed form expression). By definition, this ``coherence timescale'' is less than or equal to the total observation time $\Tobs$ over which the signal is expected to last (e.g. $\Tcoh\lesssim \Tobs\lesssim 10^{4}$\,s for the type of signals of interest in the context of GRB afterglow plateaus).

Since the signal is quasi-periodic, we can define an SFT baseline $\baseline\leq\Tcoh$ such that, within the baseline, all of the signal power is concentrated in a single SFT bin.   More specifically, around each time $T_I$ we can approximate the signal received by the detector in the time interval $T_I-\frac{\baseline}{2}\lesssim t \lesssim T_I+\frac{\baseline}{2}$, as:
\begin{multline}
h(t)\approx  h_0(T_I)\A_+F_+\cos(\Phi(T_I)+2\pi f(T_I)(t-T_I))+\\
h_0(T_I)\A_{\times}F_{\times}\sin(\Phi(T_I)+2\pi f(T_I) (t-T_I)),
\label{segnale}
\end{multline}
where $\A_{+},~\A_{\times}$ are amplitude factors dependent on the physical system's inclination angle $\iota$ (for on-axis GRBs, $\iota$ is the angle between the jet axis and the line of sight):
\begin{align}
\A_+&=\frac{1+\cos^2\iota}{2},\\
\A_{\times}&=\cos\iota,
\end{align}
and $F_{+},~F_{\times}$ are the antenna factors that quantify the detector's
sensitivity to each polarization state. Note that for triggered searches
targeting GRBs (as is the case in Sec. \ref{sec:plateau}), the line of sight
is expected to be nearly aligned with the jet axis,\footnote{That is, the line of
sight is within the jet-opening angle, which is expected to be of the order $5-20$\,deg for long GRBs \cite{Frail2001,Panaitescu2001}.} thus $\iota\approx0$ and $\A_+\approx\A_\times\approx1$.

In order for the approximation in Eq. \eqref{segnale} to be valid, the following conditions should be satisfied:
\begin{enumerate}
\item $\Tobs \lesssim 10^4$\,s so that, for a given GW detector, $F_+$ and $F_{\times}$ can be treated as constants as a function of time (see e.g. \cite{Dhurandhar2008}).
\item If $\dot{f}(t)$ is the time derivative of the signal frequency at a given time $t$, then the effects of $\dot{f}(t)$ on the signal phase should be negligible during the time interval $\baseline$. Using the quarter-cycle criterion, this leads to $2\pi |\dot{f}(T_I)|\left(\frac{\baseline}{2}\right)^2 < \frac{\pi}{2}$. Thus, $\baseline<1/\sqrt{|\dot{f}(T_I)|}$.
\item $\baseline$ is small enough that $h_0(t)\approx h_0(T_I)$ (constant amplitude approximation) in the interval $T_I-\baseline/2 \lesssim t \lesssim T_I+\baseline/2$. We consider this condition satisfied if $\left|\dot{h}_0(T_I)\right|\baseline/h_0(T_I)\lesssim 10\%$, based on typical LIGO amplitude calibration errors ($\sim 10\%$ \citep{Abadie:2010px};  thus, any change of signal amplitude below $10\%$ is not expected to significantly affect the goodness of this approximation).
\end{enumerate}
In addition, hereafter we assume that $\baseline$ is large enough that the corresponding frequency resolution, $(\baseline)^{-1}$, still enables one to track the time-frequency evolution of the signal.

Using Eq. (\ref{eq:dft}), we can calculate the DFT of the signal in Eq. (\ref{segnale}) (see also Eq. (2.25) in \cite{Dhurandhar2008}):
\begin{multline}
\tilde{h}_I[f_k]=h_{0}(T_I)e^{i\pi f_{k,I}\baseline}\times\\ [e^{i\Phi(T_I)}\frac{\A_+F_{+,I}-i\A_{\times}F_{\times,I}}{2}\delta_{\baseline}(f_k-f_{k,I})+\\e^{-i\Phi(T_I)}\frac{\A_+F_{+,I}+i\A_{\times}F_{\times,I}}{2}\delta_{\baseline}(f_k+f_{k,I})],
\label{SFT}
\end{multline}
or, equivalently,
\begin{multline}
\tilde{h}_I[f_k]=\frac{\sqrt{\A^2_+F^2_{+,I}+\A^2_\times F^2_{\times,I}}}{2}h_{0}(T_I)e^{i\pi f_{k,I}\baseline}\\\times \left[e^{i\Phi(T_I)}e^{i\varphi_I}\delta_{\baseline}(f_k-f_{k,I})+\right.\\ \left.e^{-i\Phi(T_I)}e^{-i\varphi_I}\delta_{\baseline}(f_k+f_{k,I})\right],
\end{multline}
where we have set:
\begin{eqnarray}
\nonumber \A_+F_{+,I}\pm i\A_{\times}F_{\times,I}=\sqrt{\A^2_+F^2_{+,I}+\A^2_\times F^2_{\times,I}}e^{\mp i\varphi_I},\\
\end{eqnarray}
and
\begin{equation}
\varphi_{I}=\arctan(-\A_\times F_{\times,I}/\A_+F_{+,I}).
\label{eq:complex-phase}
\end{equation}
Note that, while in our limit of intermediate-duration GW signals the antenna response from one detector can be considered constant over the observed duration of the signal, for the multiple detector case the antenna responses refer to the specific GW detector from whose output the $I$-th SFT is taken.


\section{The cross-correlation statistic}
\label{sec:xcorr}
Following \cite{Dhurandhar2008}, we define the raw cross-correlation statistic as:
\begin{equation}
\mathcal{Y}_{IJ}=\frac{\tilde{x}^*_I[f_{k,I}]\tilde{x}_J[f_{k',J}]}{\Delta T^2_{SFT}},
\label{eq:raw-cross-corr}
\end{equation}
where the frequency $f_{k,I}$ is the frequency at which all of the signal power is concentrated during the $I^\mathrm{th}$ time interval (see Eq. (\ref{SFT})), and is related to the frequency $f_{k',J}$ at which all of the signal power is concentrated during the $J^\mathrm{th}$ time interval via the relation:
\begin{equation}
f_{k',J} = f_{k,I}-\Delta f_{IJ}.
\end{equation}
In the above relation, $\Delta f_{IJ}$ is the frequency difference predicted by the model's time-frequency evolution (in this analysis the signal time-frequency evolution is assumed to be known to some level of accuracy; see Section \ref{signalsft}). Note that, because for any $I$-th SFT the associated frequency bin $k$ is fixed by the model's predictions, we omit the indexes $k,~k'$ from $\y_{IJ}$ for simplicity.

For a signal embedded in stationary Gaussian noise with zero mean, the $\{\mathcal{Y}_{IJ}\}$
are themselves random variables with mean and variance given by
\begin{align}
\mu_{IJ}&=h_{0}(T_I)h_{0}(T_J)\tilde{\cal G}_{IJ},\label{eq:mu}\\
\sigma^2_{IJ} &= \frac{1}{4\baseline^2}S_n[f_{k,I}]S_n[f_{k',J}],
\label{eq:sigma}
\end{align}
where we have used Eqs. \eqref{eq:noisepsd} and \eqref{SFT}, and the fact that:
\begin{equation}
\tilde{h}^*_I[f_k]\tilde{h}_J[f_k+\Delta f_{IJ}] = h_0(T_I)h_0(T_J)\G_{IJ} \delta_{\baseline}^2(f_k-f_{k,I}).
\label{hh}
\end{equation}
In the above equations, $\G_{IJ}$ is the signal cross-correlation function, defined here as
\begin{eqnarray}
\nonumber \G_{IJ} = \frac{\sqrt{\A^2_+F^2_{+,I}+\A^2_\times F^2_{\times,I}}}{2}\frac{\sqrt{\A^2_+F^2_{+,J}+\A^2_\times F^2_{\times,J}}}{2}e^{-i\Delta \theta_{IJ}},\\
\label{eq:sig-x-corr-func}
\end{eqnarray}
with  $\Delta \theta_{IJ}=\theta_I-\theta_J=\pi\Delta f_{IJ} \baseline+\Delta\Phi_{IJ}+\Delta\varphi_{IJ}$. In general, the subscripts $(I),~(J)$ in the antenna responses refer to the specific GW detector from whose output the $I$-th (or $J$-th) SFT is taken. Indeed, in the definition of the  $\{\mathcal{Y}_{IJ}\}$, there is total freedom to correlate pairs from one single detector or  from an arbitrary number of detectors.

Note that the $e^{-i\pi\Delta f_{IJ} \baseline}$ term that arises from $\Delta\theta_{IJ}$ in Eq. (\ref{eq:sig-x-corr-func}) is absent from the definition of the signal-cross-correlation function given in
\citep{Dhurandhar2008}. This discrepancy was first noted in \citep{Chung2011}, and is discussed there in detail. This term proves essential
to properly tracking the frequency evolution of a given signal across SFTs, so we call attention to it here.

When cross-correlation pairs are only taken from the
output of a \textit{single} detector over timescales of $\Tobs \lesssim 10^4$\,s, then $F_{+,\times,I}=F_{+,\times,J}=F_{+,\times}$. This simplifies Eq. \eqref{eq:sig-x-corr-func} considerably:
\begin{equation}
\tilde{\cal G}_{IJ}^\text{1D}=\frac{\A^2_+F^2_{+}+\A^2_\times F^2_{\times}}{4}e^{-i\pi\Delta f_{IJ} \baseline}e^{-i\Delta\Phi_{IJ}}.
\label{eq:1dxcorr}
\end{equation}
For two or more detectors, such as LIGO Hanford (H) and LIGO Livingston (L), the indexes $I$ and $J$ are free to range
over SFTs from \emph{either} detector, and so the above simplification does not generally apply (even if the antenna factors for each detector are approximately constant within the considered time interval).

Following \cite{Dhurandhar2008}, our detection statistic is then constructed as a weighted sum of the $\mathcal{Y}_{IJ}$
\begin{equation}
\rho = \sum_{IJ} (u_{IJ} \mathcal{Y}_{IJ}+u^*_{IJ} \mathcal{Y}^*_{IJ}),
\label{eq:rho}
\end{equation}
with nearly optimal weights\footnote{Strictly speaking, these weights are only optimal when self-pairs are excluded, as in \citep{Dhurandhar2008}.
For sufficiently small amplitude signals, these weights remain optimal, to first order,
even when self-pairs are considered. For situations where this may not be the case, we refer the reader to the discussion in the Appendix of \citep{Dhurandhar2008}.}
\begin{equation}
u_{IJ}=\frac{\tilde{\cal G}^{*}_{IJ}}{\sigma^2_{IJ}}.
\label{eq:weights}
\end{equation}

For stationary Gaussian distributed white noise (see Eq. \ref{eq:sigma}), $\sigma_{IJ}$ does not depend on frequency nor on time, but it might still depend on the detector. Thus:
\begin{equation}
\sigma_{IJ}^2= \frac{1}{4\baseline^2}S^2_n,
\label{eq:1-det-sigma}
\end{equation}
for $IJ$ pairs from a single detector (or identical detectors), or:
\begin{equation}
\sigma_{IJ}^2= \frac{1}{4\baseline^2}S^{H}_nS^{L}_n,
\label{eq:2-det-sigma}
\end{equation}
for e.g. a LIGO Hanford-Livingston $IJ$ pair. Thus, using the above equations and Eq. (\ref{eq:sig-x-corr-func}), we have in general:
\begin{eqnarray}
\nonumber u_{IJ} =  \frac{\sqrt{(\A^2_+F^2_{+,I}+\A^2_\times F^2_{\times,I})(\A^2_+F^2_{+,J}+\A^2_\times F^2_{\times,J})}}{\baseline^{-2} e^{-i\Delta\theta_{IJ}} S_n[f_{k,I}]S_n[f_{k,J}]}, \\\label{eq:weight-magnitude}
\end{eqnarray}
where, again, the antenna responses and detector's noise refer to the specific GW detector from whose output the $I$-th (or $J$-th) SFT is taken.

As we describe in more detail in what follows, the mean and variance of $\rho$, as well its statistical distribution, depend on the choice of which SFT pairs are cross-correlated. Because of the freedom in choosing which data-segment pairs to correlate, we can naturally consider one single detector or an arbitrary number of detectors (with no need to modify our statistic), and we can work in one of the following limits \citep{Dhurandhar2008}:
\begin{enumerate}
\item We can choose to correlate only data segments taken from distinct detectors at the same times (after correcting for the light travel time between different detectors; Section \ref{Sec:stochastic}). This limit is analogous in spirit to the methods of stochastic GW searches, such as \cite{Flanagan:1993ix, Allen:1997ad, Cornish:2001hg, Ballmer:2005uw}, and we hence refer to it as the ``stochastic limit''. In this case, the computational cost of the search is small and the search is very robust against signal uncertainties. But the sensitivity is the poorest, as is the resolution in parameter space.
\item At the other extreme, we can correlate all possible SFT segments (Section \ref{Sec:matched}): This (nearly) corresponds to a full matched filter statistic described for coalescing compact binaries and continuous waves in e.g. \cite{Finn:1992wt, Brady1998, Jar1998, 2010CQGra..27s4016A}. The parameter space resolution becomes very fine and while this is ideally the most sensitive method, is it also the most computationally expensive (prohibitive for wide parameter space searches) and the least robust against signal uncertainties.
\item In intermediate regimes, we can correlate data segments separated by a
maximum coherence time $\Tcoh\lesssim \Tobs$ (Section \ref{sec:semi-coh}).
This ``semi-coherent'' approach is similar to several methods used for
continuous waves \cite{Brady2000, Krishnan:2004sv, Dergachev2005,
Dergachev:2011pd, Goetz:2015xva} (though on signal timescales much longer than what considered in this work). Because in this limit the sensitivity and robustness of the search can be tuned to the expected accuracy of a given model, this is the regime of greatest interest to us.
\item Finally, one can consider all pairs except self-correlations. This was the main focus of the analysis presented in \citep{Dhurandhar2008} (see their Section IV). Here, we do not focus on this limit because we consider it a special case of the ones above (with no particular advantages for the detection of the type of signals considered in our study and with some complications added to the statistical properties of $\rho$). However, in what follows, we do discuss the main differences of (1)-(3) above with respect to this case  (see also Section IV of \citep{Dhurandhar2008}).
\end{enumerate}

In discussing the above limits, it is useful to note that we can re-write Eq. (\ref{eq:rho}) in terms of
Eq. \eqref{eq:raw-cross-corr} as:
\begin{equation}
\rho = \frac{1}{\baseline^2}\sum_{IJ} {u_{IJ} \tilde{x}^*_I[f_{k,I}]\tilde{x}_J[f_{k',J}]+u^*_{IJ} \tilde{x}_I[f_{k,I}]\tilde{x}^*_J[f_{k',J}]},
\label{eq:rho:2}
\end{equation}
which is equivalent to:
\begin{equation}
\rho = \frac{2}{\baseline^2}\sum_{IJ} \Re\left\{{u_{IJ} \tilde{x}^*_I[f_{k,I}]\tilde{x}_J[f_{k',J}]}\right\}.
\label{eq:rho:3}
\end{equation}


\subsection{Stochastic limit (independent pairs only)}
\label{Sec:stochastic}
Consider the output of two different detectors, $\x^{H}$ and $\x^{L}$. Each detector's output can be divided into $\Tobs/\baseline=\Nsft$ segments. Of the $(2\Nsft)^2$ possible SFT pairs that can contribute to $\rho$ we correlate only pairs of SFTs from \textit{different} detectors at the same time (after correcting for the light travel time between detectors), so that $\Npairs=\Nsft$. In this limit, Eq. \eqref{eq:rho:2} becomes:
\begin{equation}
\rho = 2\sum_{I} \Re\left\{u_{II} {\cal Y}_{II}\right\},
\end{equation}
where the weights are described by e.g. Eq. \eqref{eq:2-det-sigma}. Written explicitly, this becomes
\begin{equation}
\rho =\frac{2}{\baseline^2}\sum_{I} \Re\left\{{u_{II} \tilde{x}_I^{*\text{H}}[f_{k,I}]\tilde{x}_I^\text{L}[f_{k',I}]}\right\},
\end{equation}
i.e., a weighted sum of completely independent random variables that are each the product of two Gaussian variables with mean and variance given by Eqs. \eqref{eq:mu} and \eqref{eq:sigma}.  Thus, $\rho$ converges to a Gaussian distribution (by the Central Limit Theorem) with mean (see Eqs. (\ref{eq:mu}), (\ref{eq:sig-x-corr-func}),  (\ref{eq:weight-magnitude}), and \cite{Dhurandhar2008}) and variance (see also Eq. (\ref{eq:sigma}) and \citet{Dhurandhar2008}):
\begin{widetext}
\begin{eqnarray}
\mu_{\rho}=(\A^2_+F^2_{+,H}+\A^2_\times F^2_{\times,H})(\A^2_+F^2_{+,L}+\A^2_\times F^2_{\times,L}) \frac{\baseline^2}{2}\sum_{I} \frac{h_0^2(T_I)}{{S^{H}_n[f_{k,I}]S^{L}_n[f_{k,I}]}},\label{eq:stochmean}\\
\sigma^2_{\rho} =(\A^2_+F^2_{+,H}+\A^2_\times F^2_{\times,H})(\A^2_+F^2_{+,L}+\A^2_\times F^2_{\times,L})\frac{\baseline^2}{2}\sum_{I}\frac{1}{S^{H}_n[f_{k,I}]S^{L}_n[f_{k,I}]}.
\label{eq:stochmeanvar}
\end{eqnarray}
\end{widetext}

The detection threshold is easily derived in terms of the Cumulative Distribution Function (CDF) of a normal distribution,
\begin{equation}
\normcdf(\rho) = \frac{1}{2}\left[2-\text{erfc}\left(\frac{\rho-\mu_{\rho}}{\mathcal{\sigma_{\rho}}\sqrt{2}}\right)\right],
\end{equation}
and its inverse (see also \cite{Dhurandhar2008}), where $\text{erfc}$ is the complementary error function. For a False Alarm Probability (FAP) $\alpha$, the associated threshold is simply $1-\alpha = \normcdf(\rho_\text{th})$, thus:
\begin{equation}
\rho_\text{th} = \sqrt{2}\sigma_{\rho} \text{erfc}^{-1}(2\alpha),
\end{equation}
where we have used the fact that the background distribution is considered in the absence of a signal ($\mu_{\rho}=0$). When a signal is present, the detection probability $\gamma$, or, equivalently, the False Dismissal Probability (FDP) $1-\gamma$, is given by $\gamma = \normcdf(\rho_\text{th})$, i.e.:
\begin{equation}
\gamma = \frac{1}{2}\text{erfc}\left(\frac{\rho_\text{th}-\mu_{\rho}}{\mathcal{\sigma_{\rho}}\sqrt{2}}\right).
\end{equation}

\begin{figure}[t]
\begin{center}
\includegraphics[width=0.5\textwidth,angle=0]{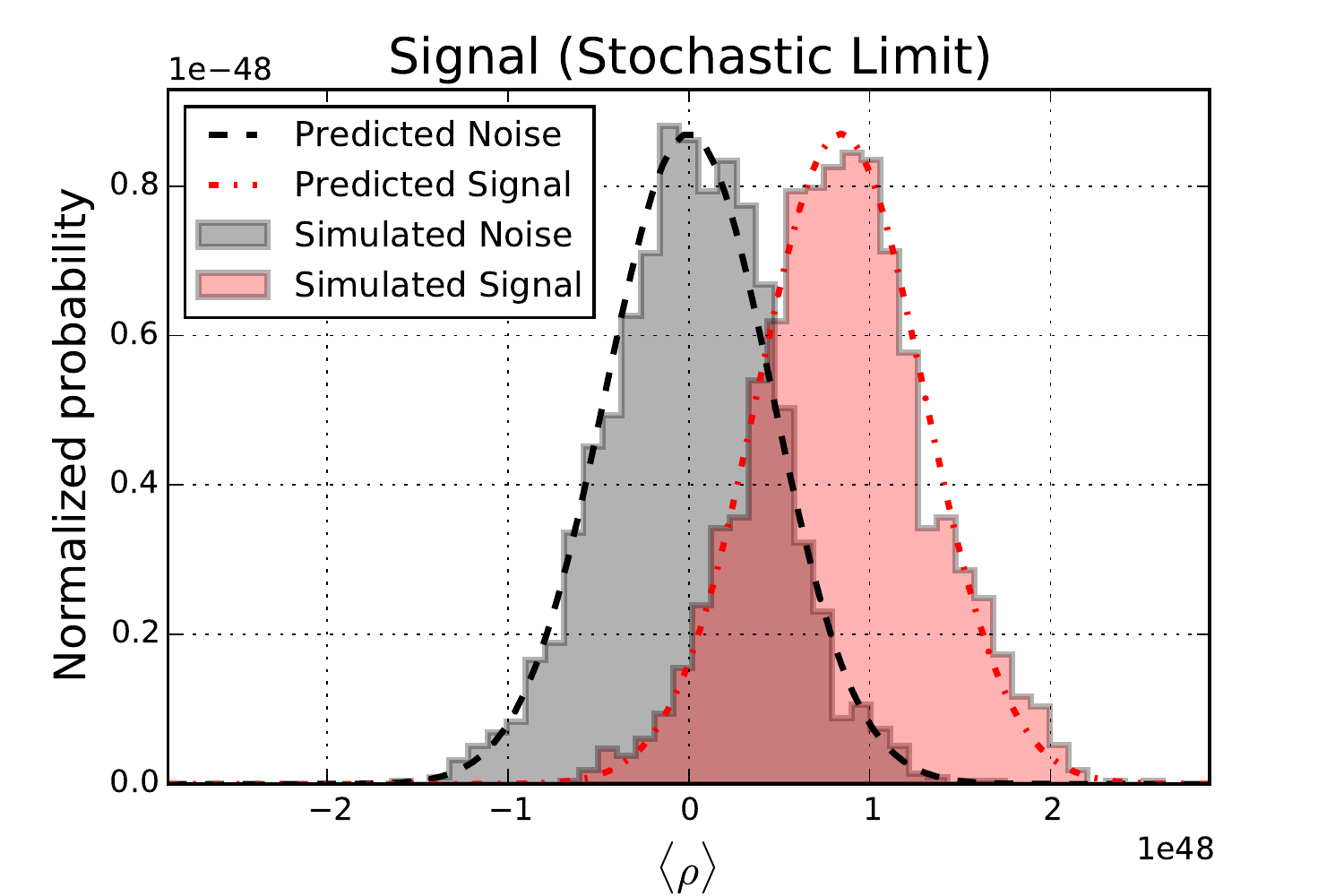}
\caption{Comparison between the simulated and predicted distribution of $\rho$ in the stochastic limit, for 2048\,s of simulated white Gaussian noise sampled at a rate of $f_s=2048\unit{Hz}$, from two detector's outputs $x^{H}[t],~x^{L}[t]$. We have used an SFT baseline of $\baseline=2\unit{s}$ and, for simplicity, we assumed two idealized, colocated, and optimally oriented detectors with aLIGO-equivalent PSDs $S_n\approx1.75\times10^{-47}~\rm{Hz}^{-1}$, see Fig. \ref{fig:noise-approximation}. The simulated signal is a line of constant frequency $f_0=128~\text{Hz}$ and constant amplitude $h_0 \approx 10^{-24}$.}
\label{fig:stochastic}
\end{center}
\end{figure}

\begin{figure}[htbp]
\begin{center}
\includegraphics[width=0.5\textwidth,angle=0]{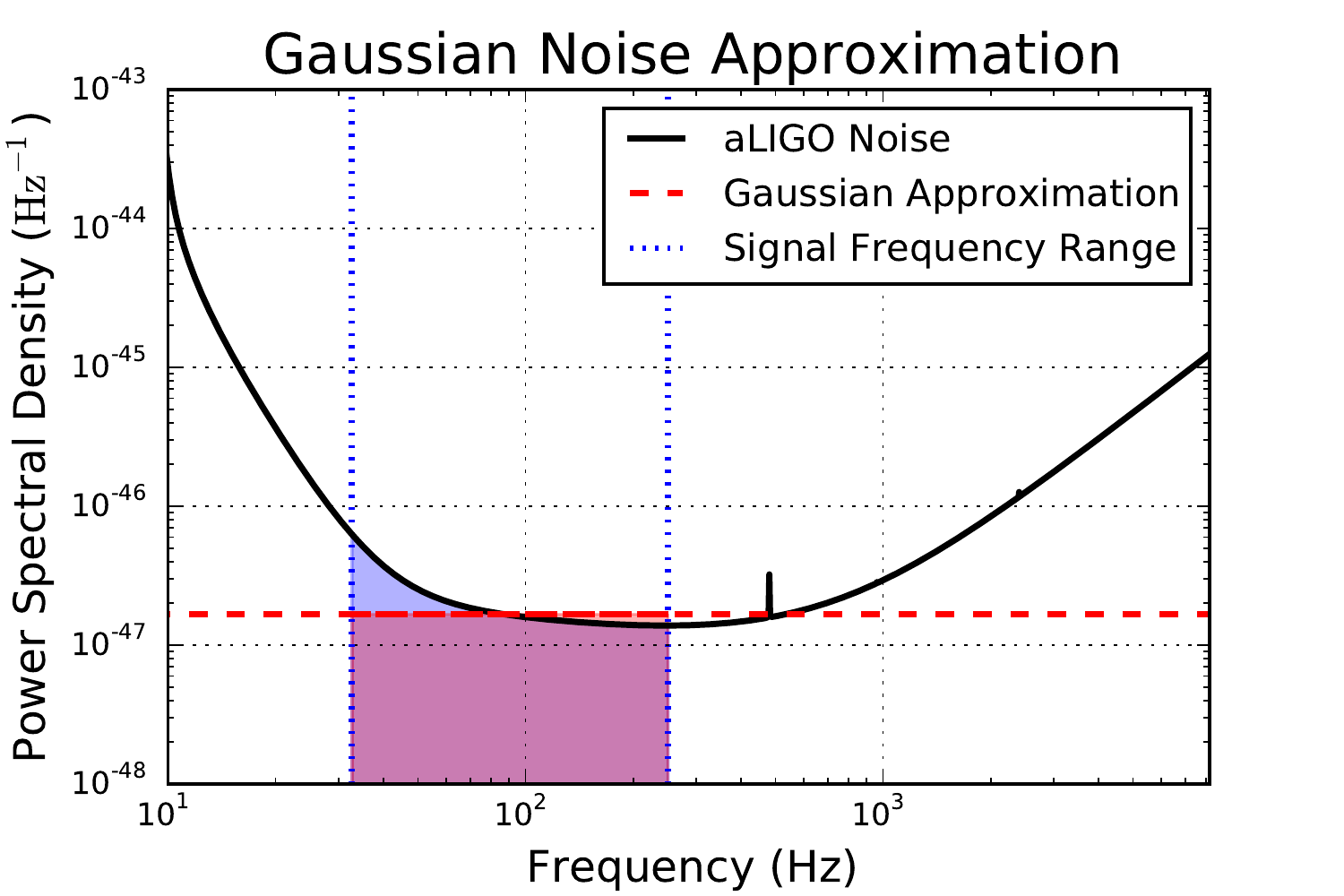}
\caption{Method for generating Gaussian-distributed noise with aLIGO PSD. A frequency range is defined with respect to the maximum and minimum value of an injected signal's frequency (blue dotted lines). The constant Gaussian PSD $S_n$ (red dashed line) is calculated so that the area underneath it (shaded red) is equal to the area underneath the aLIGO PSD (shaded blue). For signals of constant frequency $f_0$ (e.g. Figs. \ref{fig:stochastic}, \ref{fig:all-pairs}, and \ref{fig:overlap-comparison}), the red area is taken between $0.9f_0$ and $1.1f_0$.}
\label{fig:noise-approximation}
\end{center}
\end{figure}

Thus, the detectability condition reads:
\begin{equation}
\frac{\mu_\rho}{\sigma_\rho}\gtrsim \sqrt{2}{\cal S},
\end{equation}
where $\mathcal{S} = \text{erfc}^{-1}(2\alpha)-\text{erfc}^{-1}(2\gamma)$.  In
the case of white Gaussian noise, using Eqs. (\ref{eq:stochmean})--(\ref{eq:stochmeanvar}), the detectability condition implies:
\begin{equation}
  \hr\gtrsim \frac{\sqrt{2} {\cal S}^{1/2}\baseline^{-1/2}\Nsft^{-1/4}(S^{H}_nS^{L}_n)^{1/4}}{[(\A^2_+F^2_{+,H}+\A^2_\times F^2_{\times,H})(\A^2_+F^2_{+,L}+\A^2_\times F^2_{\times,L})]^{1/4} },
  \label{rmscomplicated}
\end{equation}
which generalizes Eq. (4.15) in \citet{Dhurandhar2008} to the case of a non-constant signal amplitude for which (see also Eq. \eqref{segnale}):
\begin{equation}
\hr=\sqrt{\langle h_0^2(T_I)\rangle_{I}}=\sqrt{\frac{\sum_{I} h_0^2(T_I)}{\Nsft}}.
\label{rmshdef}
\end{equation}

In Fig. \ref{fig:stochastic} we show the distribution of $\rho$ in the absence of a signal for simulated Gaussian white noise, and in the presence of a GW signal of constant amplitude $h_0$ and constant frequency $f_0$. (A signal with constant frequency represents the simplest time-frequency evolution to which the technique here presented can be applied and is particularly useful for illustrative purposes.)

We stress that the independence of the pairs that are added in $\rho$
is \emph{essential} for the validity of the conclusion regarding the Gaussianity of $\rho$, and for the validity of Eqs. (\ref{rmscomplicated})-(\ref{rmshdef}). While pairs
are truly independent in the stochastic limit analyzed in this Section, this
is not strictly true for the combination of pairs considered in Section IV of
\citep{Dhurandhar2008} ($\rho$ includes all possible pairs but self ones - see
also case 4 in Section \ref{sec:xcorr}) and in the Appendix of
\citep{Dhurandhar2008}  ($\rho$ includes all possible SFT pairs - see also
case 2 in Section \ref{sec:xcorr}). In these cases, $\rho$ is a sum of
products that are \textit{not} all independent. Thus, while the expressions
for the mean and variance of $\rho$ presented in Section IV of
\citep{Dhurandhar2008} (or, equivalently, Eqs. (\ref{eq:stochmean}) and
(\ref{eq:stochmeanvar}) here) remain valid, we caution the reader that the
\textit{lack of independence affects the shape of the background
distribution}, and in some limits, results in a distribution that
\textit{cannot} be reduced to a Gaussian. Thus, the detection
threshold needs to be modified accordingly. Some brief discussion of these
corrections to \citep{Dhurandhar2008} is also presented in Appendix B of
\cite{Whelan2015}. In what follows, our in depth discussion of cases 2--3 (Section \ref{sec:xcorr}) shows explicitly that the corrections to \citep{Dhurandhar2008} are crucial for the detection of the family of intermediate-duration GW signals that we target in this analysis.


\subsection{Matched filter limit (all pairs)}
\label{Sec:matched}

In this limit, we choose to correlated all possible SFT segments (from one or multiple detectors). Starting from Eq. (\ref{eq:rho:3}), we replace the weights with their explicit form given by Eq. \eqref{eq:weight-magnitude},
\begin{widetext}
\begin{eqnarray}
\rho=2\Re\left[\sum\limits_{I,J}^{\Npairs}\frac{\sqrt{(\A^2_+F^2_{+,I}+\A^2_\times F^2_{\times,I})}}{S_n[f_{k,I}]}\frac{\sqrt{(\A^2_+F^2_{+,J}+\A^2_\times F^2_{\times,J})}}{S_n[f_{k,J}]}\x^*_I[f_{k,I}]\x_J[f_{k,J}]e^{i\Delta\theta_{IJ}}\right],\label{eq:eq-chi2-part1}
\end{eqnarray}
\end{widetext}
where $\Npairs=\Nsft^2$ and $\Nsft=\Ndet\Tobs/\baseline$, with $\Ndet$ being the number of detectors from which data are taken. Under the change of variable
\begin{equation}
\x_I'[f_{k,I}]=\frac{\sqrt{(\A^2_+F^2_{+,I}+\A^2_\times F^2_{\times,I})}}{S_n[f_{k,I}]} \x_I[f_{k,I}]e^{-i\theta_I},
\label{eq:change-of-variable}
\end{equation}
Eq. \eqref{eq:eq-chi2-part1} simplifies to:
\begin{equation}
\rho = 2\left\{\sum\limits_I^{\Nsft}\left|\tilde{x}'_I[f_{k,I}]\right|^2+2\sum\limits_{I>J}^{\Nsft}\Re\left[{\x_I^{\prime*}}[f_{k,I}]\x_J'[f_{k',J}]\right]\right\}.
\end{equation}
It then follows that,
\begin{equation}
\rho = 2\left|\sum\limits_I^{\Nsft} \x_I'[f_{k,I}]\right|^2.
\label{eq:eq-chi2-part2}
\end{equation}
Or alternatively,
\begin{equation}
\rho = 2\left[\left(\sum\limits_I^{\Nsft} \Re(\x_I'[f_{k,I}])\right)^2+\left(\sum\limits_I^{\Nsft} \Im(\x_I'[f_{k,I}])\right)^2\right].
\label{eq:eq-chi2-part3}
\end{equation}

\begin{figure}
\begin{center}
\includegraphics[width=0.5\textwidth,angle=0]{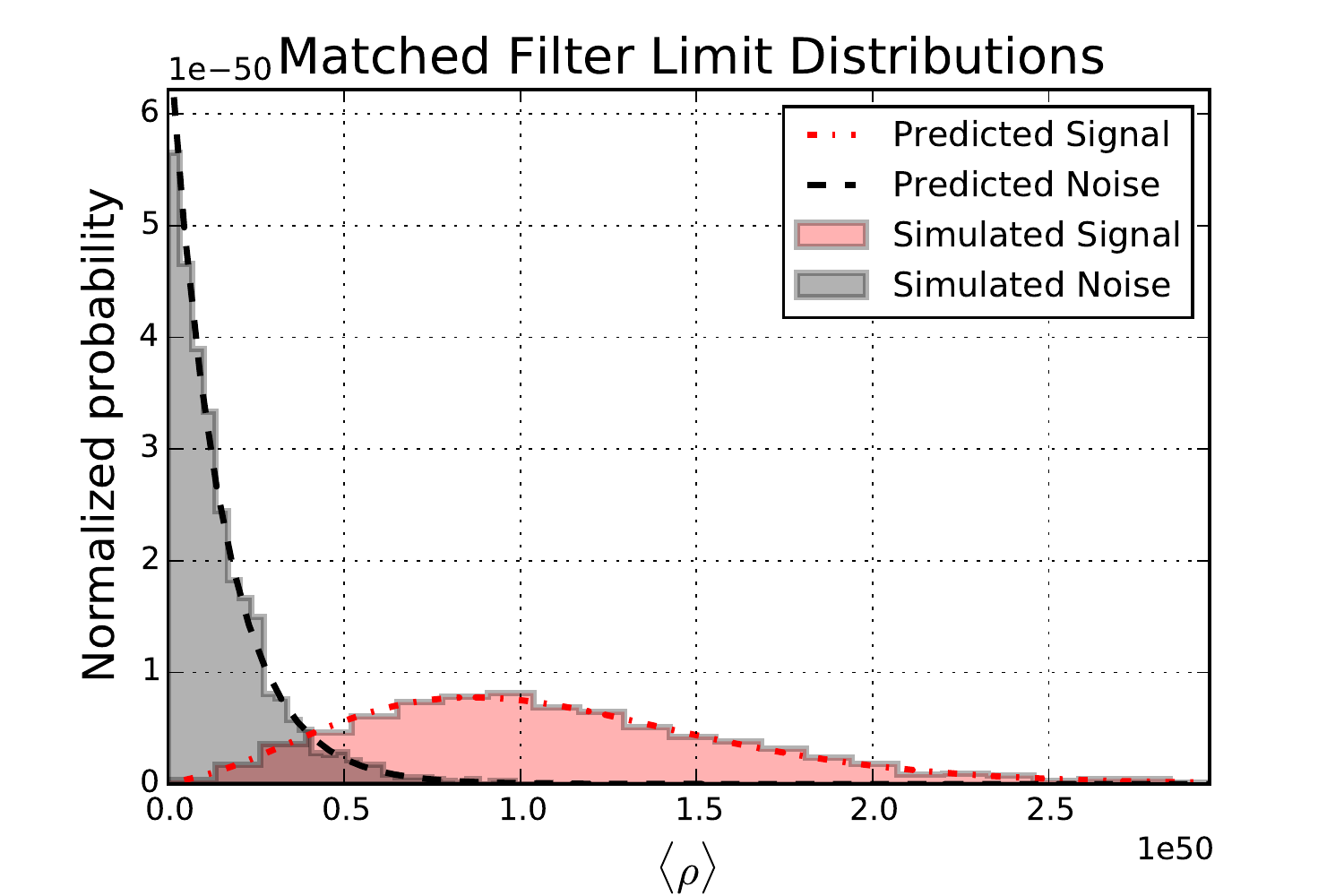}
\caption{We simulate $2048\unit{s}$ of white Gaussian noise for a single optimally-oriented detector with aLIGO-equivalent noise PSD given by $S_n\approx1.75\times10^{-47}~\rm{Hz}^{-1}$. We used a sampling frequency of $f_s=2048\unit{Hz}$ and an SFT baseline of $\baseline=2\unit{s}$. All possible pairs are included in the cross-correlation statistic $\rho$ which is thus distributed as a scaled $\chi^2$ distribution with 2 degrees of
freedom. The simulated signal was a line of constant frequency $f_0=128~\text{Hz}$ and constant amplitude $h_0 \approx 3.30\times10^{-25}$.}
\label{fig:all-pairs}
\end{center}
\end{figure}

For stationary Gaussian noise with zero mean, $\x_I[f_k]$ follows a complex normal distribution. We note that the scaling and complex rotation applied in Eq. \eqref{eq:change-of-variable} have no effect on the shape of the distribution of the $\x'_I$ when compared to the $\x_I$ (but they do change the mean and variance of the distribution). Thus, the real and imaginary parts of $\x'_I$ are still Gaussian distributed as the $\x_I$, and so are their sums. Indeed, in the absence of a signal, the sums of the real and imaginary parts of the $\x'_I$ are Gaussian variables with zero mean and variance (see Eq. \eqref{eq:noisepsd}):
\begin{equation}
\sigma^2_{\Sigma}=\sum_I^{\Nsft}\left[\frac{\baseline(\A^2_+F^2_{+,I}+\A^2_\times F^2_{\times,I})}{4S_n[f_{k,I}]}\right].
\end{equation}
So we can re-write the expression for $\rho$ as:
\begin{multline}
\rho = \scale \times \left[\left(\frac{\sum_I^{\Nsft} \Re(\tilde{x}_I'[f_{k,I}])}{\sigma_{\Sigma}}\right)^2\right.\\
\left.+\left(\frac{\sum_I^{\Nsft}\Im(\tilde{x}_I'[f_{k,I}])}{\sigma_\Sigma}\right)^2\right],
\label{eq:rho:xhi2}
\end{multline}
which is the sum of the squares of two normally distributed variables, scaled by a factor:
\begin{eqnarray}
\nonumber \scale = 2\sigma_\Sigma^2=\sum_I^{\Nsft}\left[\frac{\baseline(\A^2_+F^2_{+,I}+\A^2_\times F^2_{\times,I})}{2S_n[f_{k,I}]}\right].\\
\label{eq:scale-factor}
\end{eqnarray}
Thus, the resulting $\rho$ statistic is distributed as a $\chi^2$ with 2 degrees of freedom (Fig. \ref{fig:all-pairs}; see also \cite{Whelan2015}).

Continuing from Eq. \eqref{eq:rho:xhi2}, in the absence of a signal, the variance of $\rho$ is simply,
\begin{align}
\sigma_{\rho}^2 &= 4\scale = 2\sum_I^{\Nsft}\left[\frac{\baseline(\A^2_+F^2_{+,I}+\A^2_\times F^2_{\times,I})}{S_n[f_{k,I}]}\right].
\label{eq:mf-variance}
\end{align}

In the presence of a signal, the distribution of $\rho$ in Eq. \eqref{eq:rho:xhi2} becomes a non-central $\chi^2$ with two degrees of freedom, $\chi^2_\textsc{nc}(2;~\lambda)$, of mean:
\begin{align}
\mu_{\rho} &= \scale(2+\lambda) \label{eq:matched-filter-mean}.
\end{align}
The  non-centrality parameter can be derived using the above relation, and noting that $\mu_{\rho}$ can be easily calculated using Eqs. \eqref{hh}, \eqref{eq:sig-x-corr-func}, and \eqref{eq:eq-chi2-part1}. This yields (see also Eq. \eqref{rmshdef} and Fig. \ref{fig:all-pairs}):
\begin{equation}
\lambda = \sum_I^{\Nsft}h^2_0(T_I)\left[\frac{\baseline(\A^2_+F^2_{+,I}+\A^2_\times F^2_{\times,I})}{S_n[f_{k,I}]}\right].
\label{eq:non-centrality-parameter}
\end{equation}
Note that in this limit the number of SFT pairs only affects the variance (and
mean) of the two Gaussian variables $\sum_I^{\Nsft} \Re(\tilde{x}'_I[f_{k,
I}])$ and $\sum_I^{\Nsft} \Im(\tilde{x}'_I[f_{k, I}])$. It does not affect the
number of degrees of freedom in $\rho$, which remains two independently of the number of SFTs. Thus, as $\Nsft$ increases, the distribution of $\rho$ does \textit{not} approach a Gaussian. This is a critical distinction to make, since it changes the (false alarm and false dismissal) thresholds of $\rho$ significantly from the ones that were adopted in the appendix of \citep{Dhurandhar2008}, where a Gaussian distribution was incorrectly assumed for $\rho$.

In the case in which all pairs come from a single detector (or from colocated, equally oriented detectors, with identical $S_n$), the variance of $\rho$ simplifies substantially to:
\begin{align}
\sigma_{\rho}^2 &= 4\scale = 2\Tobs\left[\frac{(\A^2_+F^2_{+}+\A^2_\times F^2_{\times})}{S_n}\right],
\end{align}
where we have used $\Tobs=\Nsft \baseline$. The non-centrality parameter likewise simplifies, yielding,
\begin{equation}
\lambda = \ncp,
\end{equation}
where we have used Eq. \eqref{rmshdef}.

In either case, the corresponding detection threshold for a given false alarm and detection rate is now substantially different than in the stochastic limit:
\begin{align}
\rho_\text{th}&=\scale \chicdf^{-1}(1-\alpha; 2),\\
\gamma &= \ncxcdf(\rho_\text{th}/\scale; 2, \lambda).
\end{align}
The CDF for the $\chi^2(2)$ is known in closed form (and is even invertible), while the CDF for the non-central case can be calculated numerically, with results as shown in Fig. \ref{fig:sensitivity}.

In this limit, the sensitivity approaches that of matched filtering.
However there is one significant error in the description in
\cite{Dhurandhar2008}: the limit approached is that of filtering with an
unknown overall phase constant, which is commonly handled by summing the
squares of two matched filters a quarter cycle out of phase with each
other---e.g., \cite{Owen1996}.
Hence the resulting statistic is distributed as a $\chi^2$ with 2 degrees of
freedom rather than a Gaussian.
Under idealized circumstances, this reduces the sensitivity by approximately
13\% with respect to a Gaussian distribution (with FAP=0.1\% and FDP=50\%).


\subsection{Semi-coherent regime}
\label{sec:semi-coh}

\begin{figure}
\begin{center}
\includegraphics[width=0.5\textwidth,angle=0]{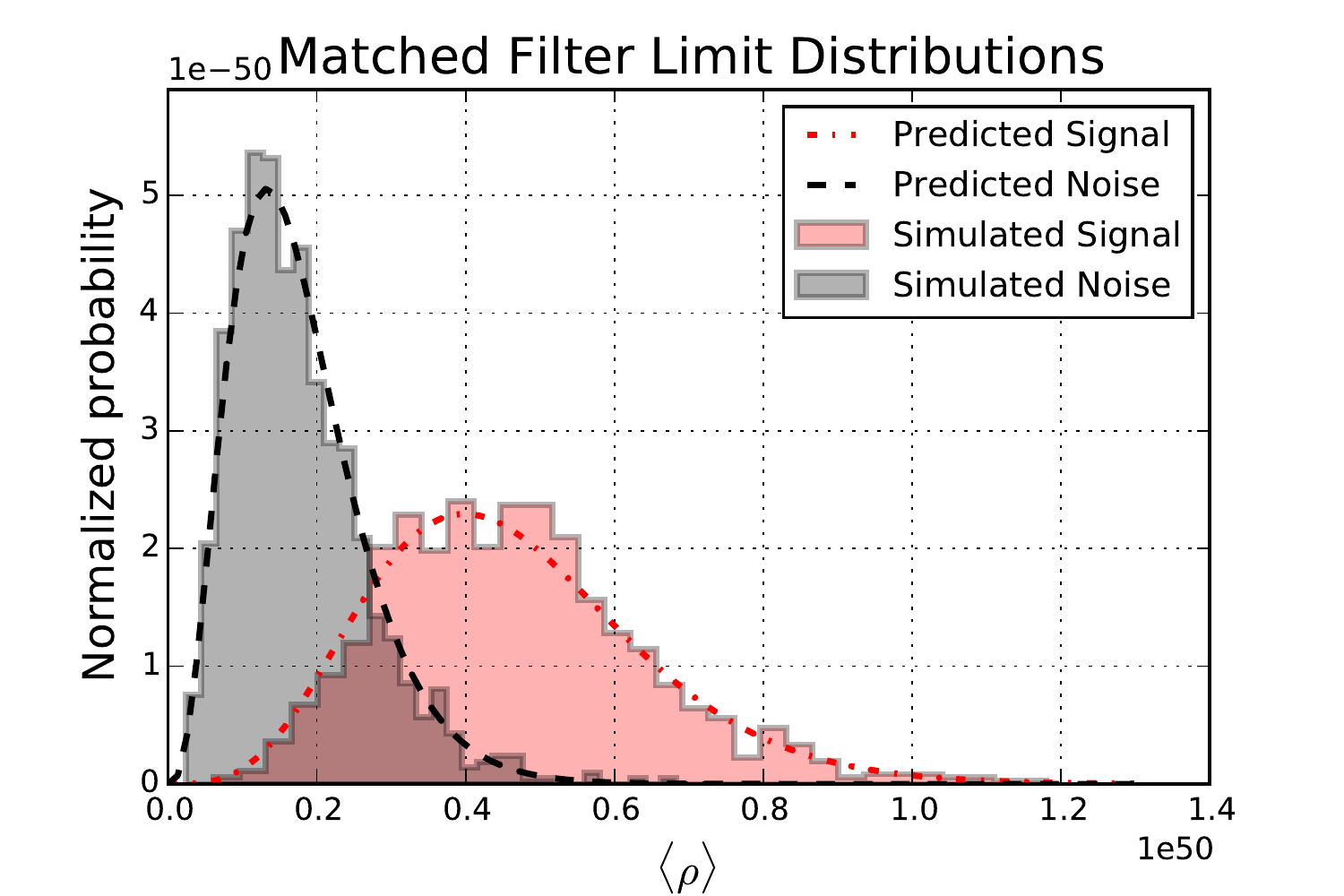}
\caption{Comparison between the simulated and predicted distribution of $\rho$ in the semi-coherent limit, for 1024\,s of simulated white Gaussian noise sampled at a rate of $f_s=2048\unit{Hz}$, from one detector's output $x(t)$. We have used an SFT baseline of $\baseline=2\,\rm{s}$ and we assumed an optimally oriented detector with PSD $S_n\approx 1.91\times10^{-47}\,\text{Hz}^{-1}$. The coherence time is $\Tcoh=256$\,s for a total of $\Ncoh=4$ coherent segments. The simulated signal was a line of constant frequency $f_0=128~\text{Hz}$ and constant amplitude $h_0 \approx 8.47\times10^{-25}$.}
\label{fig:overlap-comparison}
\end{center}
\end{figure}

As discussed in Section \ref{sec:motivation}, the semi-coherent regime is the most relevant for an astrophysically motivated search where the expected GW signal is known to limited accuracy.
In this regime, the total observation time $\Tobs$ is broken up into $\Ncoh$ coherent segments, each of duration $\Tcoh$.
The coherence time ($\Tcoh$) is once again defined as the length of time wherein the signal is expected to maintain phase coherence (and therefore good agreement) with
the model predictions. All possible SFT-pairs within each coherent time segment are cross-correlated, and the results for each segment are then
combined incoherently.\footnote{An alternative, but equivalent, description is to define a ``coherence window'' of duration $\Tcoh$ which is
then stepped across the SFT according to a given spacing criterion. All segments in each step are cross-correlated then combined
incoherently.}

In order for the resulting sum of $\chi^2(2)$ distributed variables to add to
a $\chi^2(2\Ncoh)$ distributed detection statistic, it is essential that all
coherent segments have identical scale factors.\footnote{In general, for random
$\chi^2$ variables $X_i$, their linear combination $Y=\sum_i \mathcal{C}_iX_i$
is itself a $\chi^2$ variable if and only if the scale coefficients
$\mathcal{C}_i$ are identical (or 0). However, if the normalized coefficients
$\mathcal{C}_i/\langle\mathcal{C}_i\rangle$ are close to unity, $Y$ is
reasonably \textit{approximated by} a $\chi^2$ distribution.} This condition is satisfied
for a detector network of arbitrary size only if the detectors have similar antenna factors
for the given sky location of the event,
and each detector has (stationary) white Gaussian noise (although the frequency
independent $S_n$ of the detectors need not be identical).  In the case of colored
noise, the scale factors will vary between coherence segments (since the
frequency of the signal is evolving with time, which causes $S_n[f_{k,I}]$ to
change from segment to segment). Thus, in the presence of colored noise, whitening the data over the signal bandwidth prior to analysis is desirable. 

Changes in the antenna factors $F_+,~F_\times$ over the duration
of a signal in a non-idealized search i.e., deviations from assumption 1 in Section \ref{signalsft}, can also affect the statistic. For the GRB X-ray 
plateaus of interest to Section \ref{sec:plateau}, $>$50\% of events with sufficiently shallow plateau decays\footnote{We consider specifically
Type IIa GRBs as described in \cite{Margutti2013} with plateau power law decay 
indexes of magnitudes $\lesssim$0.5.} have plateau durations $\lesssim 10^4$\,s \cite{Margutti2013}. 
For circularly-polarized signals of this duration, we tentatively estimate that time-varying antenna factors will cause fluctuations of 
$\approx$15\% in amplitude sensitivity, comparable to LIGO amplitude calibration uncertainties \cite{Abadie:2010px}. We leave to future work 
a more in depth examination of deviations from this assumption.

When all coherent segments have identical scale factors, $\rho$ is an incoherent sum of $\Ncoh$ \emph{independent} variables, each distributed as a scaled $\chi^2(2)$ distribution.
The scale parameter for each coherent segment is given by Eq. \eqref{eq:scale-factor} but now with $\Nsft=\Tcoh/\baseline$, so that:
\begin{equation}
\scale^\text{SC} = \frac{\scale}{\Ncoh}.
\end{equation}
The variance of the semi-coherent $\rho$ then reads:
\begin{equation}
\sigma_{\rho,\text{SC}}^2 = 2\scale^\text{SC}(2\Ncoh) = 4\scale,
\end{equation}
which is identical to the variance in matched-filter limit, see Eq. \eqref{eq:mf-variance}.

\begin{figure}[htbp]
\begin{center}
\includegraphics[width=0.495\textwidth]{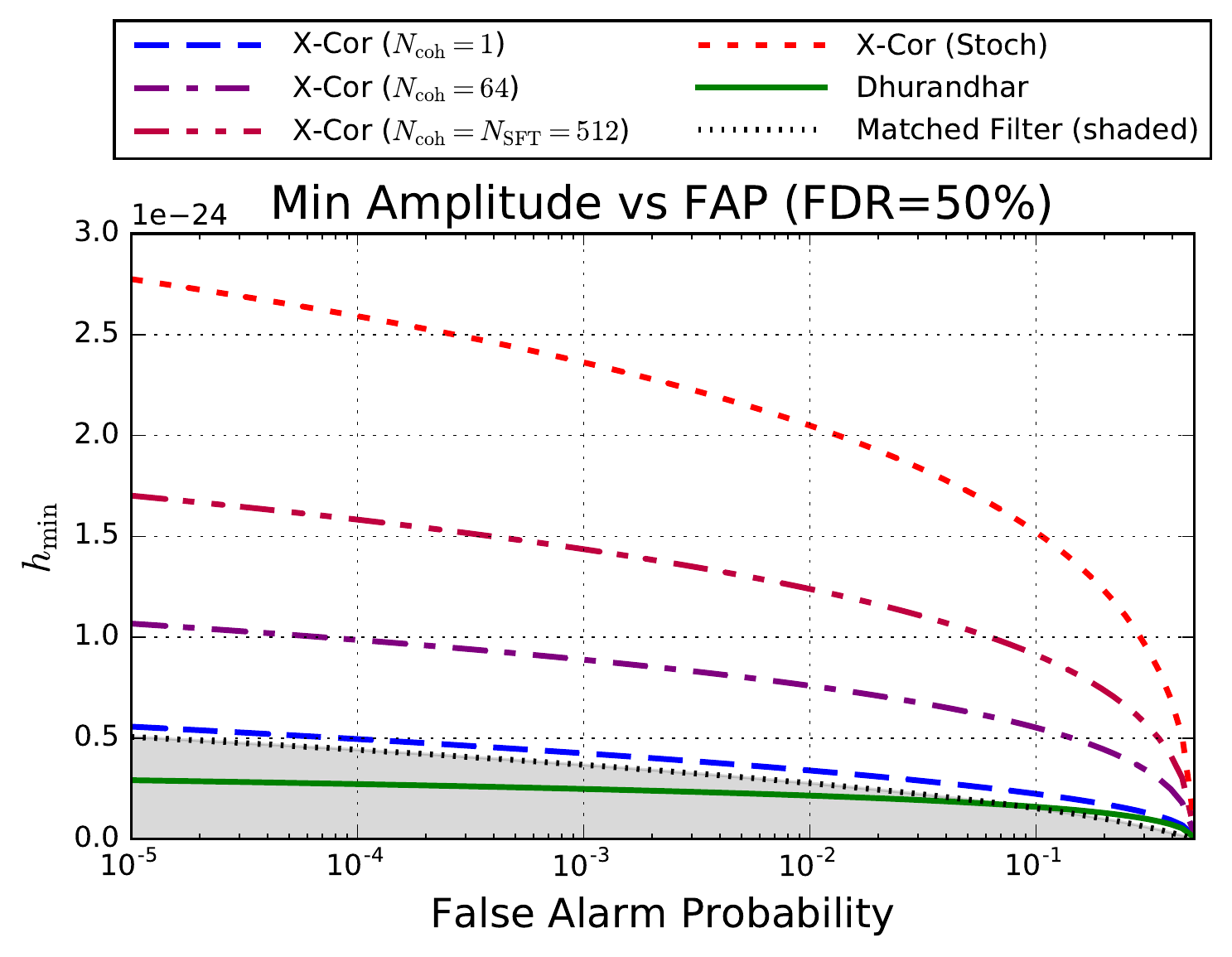}
\caption{The smallest detectable GW amplitude $h_\text{min}$ is plotted versus FAP with a set FDP of $1-\gamma = 50\%$. A matched filter with known initial phase (black dotted line, with gray shading) is the idealized optimal search, and it provides an absolute limit on the sensitivity of \emph{any} real search. Hence, the shaded gray area is forbidden. The matched-filter limit of the cross-correlation method (dashed blue line) is expected to approach (but not converge with) the black-dotted line. The semi-coherent limit (dash/dotted purple lines) becomes less sensitive for increasing $\Ncoh$, eventually approaching the stochastic limit (dotted red). The $\Ncoh=\Nsft=512$ limit of the cross-correlation method (red dash/double-dot red line) differs from the stochastic limit in that it includes self-pairs (autocorrelations).  As discussed in the text, the assumptions of Gaussian statistics and known phase constant in \citep{Dhurandhar2008} yield incorrect results as the resulting sensitivity (green solid line) does better than the optimal matched filter for sufficiently small FAP.}
\label{fig:sensitivity}
\end{center}
\end{figure}

When a signal is present, the non-centrality parameter for each coherent segment will, in general, vary from one semi-coherent chunk to the other due to the time-varying amplitude of the signal in Eq. \eqref{eq:non-centrality-parameter}. But, since the total $\lambda$ for the semi-coherent regime is additive across \textit{all} coherent segments, the \textit{total} non-centrality parameter for the semi-coherent $\rho$ likewise remains unchanged from the matched filter limit. The resulting distribution thus has mean:
\begin{equation}
\mu_{\rho,\text{SC}} = \scale^\text{SC}(2\Ncoh+\lambda).
\end{equation}
The above Equation reduces to \eqref{eq:matched-filter-mean} in the limit of $\Ncoh = 1$ (matched-filter limit). The detection threshold for a given signal will differ from the matched filter limit due to the higher number of degrees of freedom of the  $\chi^2$ distribution of the semi-coherent $\rho$, and can be calculated numerically as shown in Fig. \ref{fig:sensitivity} (along with other limits).

In the limit of large $\Ncoh$, the $\chi^2(2\Ncoh)$ distribution tends toward
a Gaussian.
Continuous wave searches using this cross-correlation technique, e.g.
\citep{Whelan2015} can have $\Ncoh$ of order $10^4$, and hence can set their
thresholds based on Gaussian statistics as assumed by \cite{Dhurandhar2008}.
However, searches for intermediate-duration GW signals (such as those of
interest to this paper) can have $\Ncoh$ smaller by 1--2 orders of magnitude,
so it is essential to correct the corresponding detection thresholds to
account for non-Gaussianity.
In particular, the Central Limit Theorem reduces the skew of the
$\chi^2(2\Ncoh)$ distribution relatively slowly as $\Ncoh$ grows.

We finally remark that, for simplicity, we have assumed coherence segments that do not overlap and no windowing function for the SFTs. For a more detailed discussion of the effects of overlapping segments and windowing, see \cite{Sundaresan2012}.

\subsection{Spectral Leakage Effects}
\label{sec:specleak}
Several  of the assumptions made in the previous Sections are expected to lead to some amount of
spectral leakage. These include the finite-time approximation of the delta function in Eq. \eqref{eq:finite-delta},
the quarter-cycle criterion, and SFT windowing effects (that is, the simplification of using a simple rectangular window). A full treatment of the effects of spectral leakage is outside the scope of this paper but we mention some of its effects
here.

\begin{figure*}
    \subfloat[SNR variation with respect to the location of $f_0$. The SNR for a line ($f(t)=f_0$) is at a maximum when $f_0$ lies at the center of one of the bins, and at a minimum when it lies on an edge. In the case of the latter, the SNR is reduced by a factor of 2, as the leakage leads to approximately half of the signal power leaking into each adjacent bin. Thus, the maximum loss in amplitude sensitivity from spectral leakage should be no more than $\sqrt{50\%}\approx71\%$ for a signal satisfying the quarter-cycle criterion (see Sec. \ref{signalsft}). The grey shaded region corresponds to the 3$\sigma$ error region resulting from 512 independent simulations. \label{fig:leakage-1-a}]{\includegraphics[width=.500\textwidth]{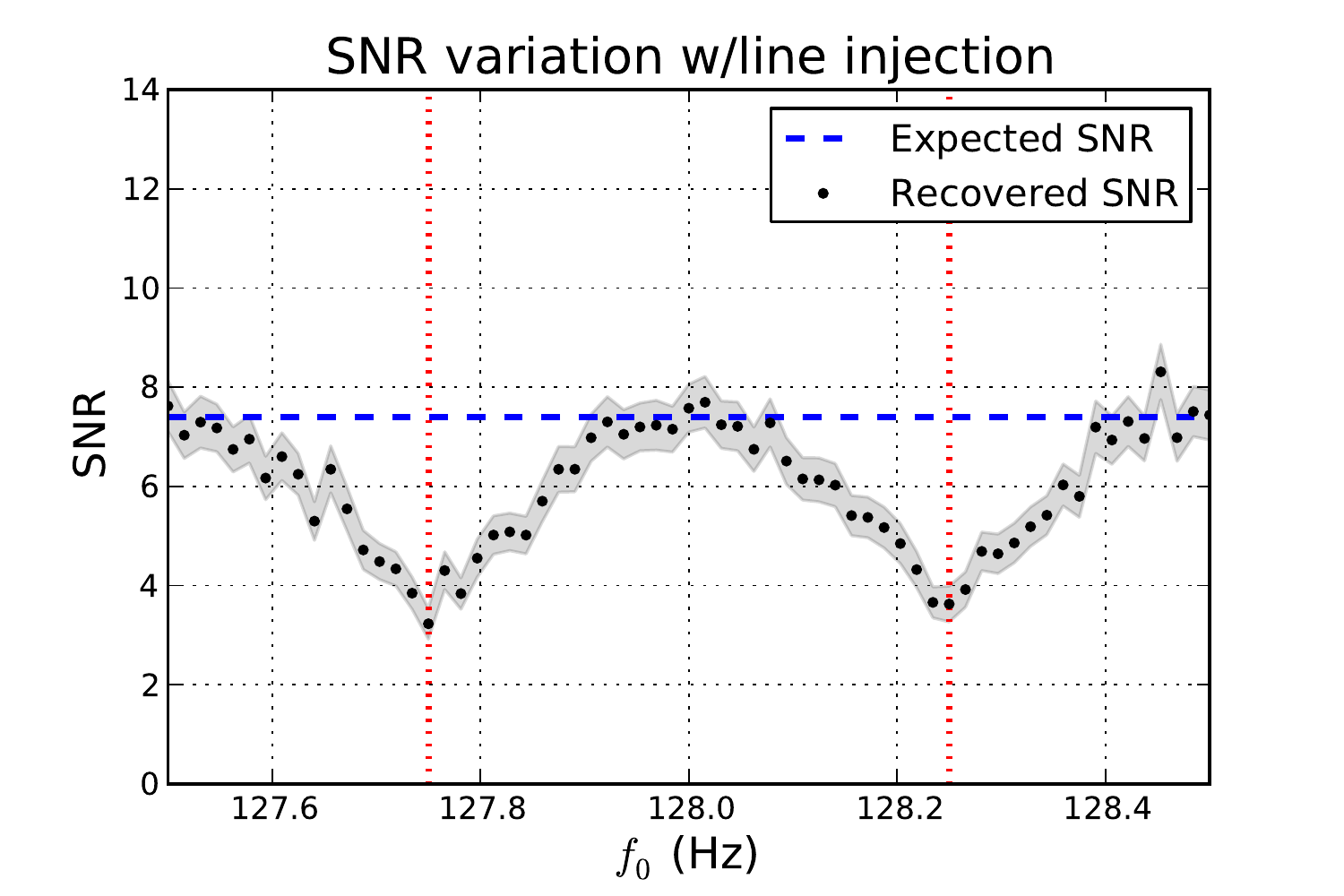}} \hfill
    \subfloat[SNR variation with respect to the location of $f_0$ for a pulsar-like evolution, $f(t)=f_0+f_1t$, where $f_1=-1/1024\,\rm{Hz}$ was chosen such that the frequency would change by one frequency bin ($1/\baseline\,\rm{Hz}$) over the entire duration of the signal ($\Tobs$), well within the quarter-cycle criterion. Here, the factor of $f_1$ guarantees that the signal never precisely corresponds to the center of the bin, which averages out to roughly 75\% of the SNR (or $\sqrt{75\%}\approx87\%$ in amplitude) for any given $f_0$ The grey shaded region corresponds to the 3$\sigma$ error region resulting from 512 independent simulations.\label{fig:leakage-1-b}]{\includegraphics[width=.49\textwidth]{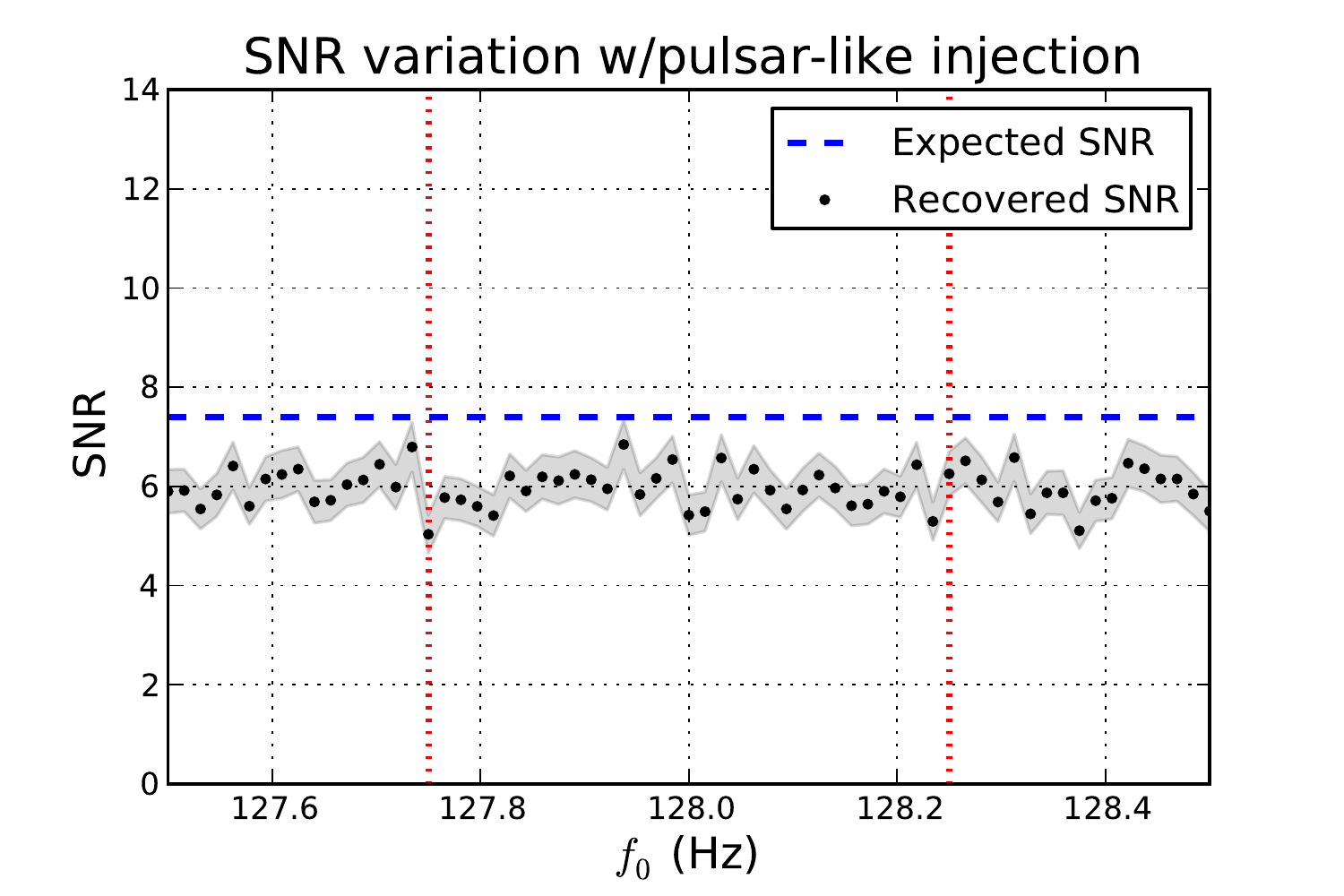}}
    \caption{The effects of spectral leakage on signals of the form $h(t)=h_0\sin\Phi(t)$ with $\Phi(t)=2\pi\int f(t)dt$ and $h_0=10^{-24}$ injected into Gaussian noise with zero mean and $S_n\approx 1.75\times10^{-47}\,\rm{Hz}^{-1}$. Two frequency evolutions are considered: a line feature of constant frequency $f(t)=f_0$ (left); a pulsar-like evolution of the form $f(t)=f_0-f_1t$ (here $f_1=1/1024\,\rm{Hz}/s$, right). The SFT baseline used to calculate the SNR is $\baseline=2$ s, resulting in SFT bin widths of $1/2$ Hz (e.g. the center of the bin located at $f_0=128.0$ Hz has edges at $128.25$ and $127.75$, red vertical dashed lines). In all cases the total duration of the signal is $\Tobs=512\,\rm{s}$.}
    \label{fig:leakage-1}
\end{figure*}

 As shown in Fig. \ref{fig:leakage-1}, spectral leakage is an issue any time the signal frequency does not precisely correspond to the center of one of the SFT
frequency bins. In the simplest case of a constant frequency periodic signal, spectral leakage can cause a reduction of up to 50\% in the SNR ($\mu_{\rho,\rm{signal}}/\sigma_{\rho,\rm{noise}}$) for the $\rho$ statistic in each of the fully coherent segments. This effect is worsened when one considers time-varying frequencies: while the quarter-cycle criterion restricts the leakage from first order terms ($\dot{f}$), higher order components of the frequency evolution ($\ddot{f},~\dddot{f},$ etc) can lead to additional leakage. The net result is that, on average, neglecting spectral leakage will result in reduced SNR that is roughly 75\% of the idealized case, see Figures \ref{fig:leakage-1-a} and \ref{fig:leakage-1-b}, and also \cite{Sundaresan2012}.

The typical solution for this problem is to introduce a windowing function for the SFT, but this is not without tradeoffs. Each windowing function
(of which there are many) has different strengths and weaknesses. The commonly used Hann window is well equipped to handle
spectral leakage and maintains good frequency resolution, but suffers in amplitude accuracy \citep{Sundaresan2012}. SFT windows must then be overlapped in an
attempt to regain some of the lost amplitude information, increasing computational cost. The Tukey window, commonly used in continuous wave searches,
is -- by contrast -- not as good at diminishing the effects of spectral
leakage but retains more of the original power. Recent work within the cross-correlation framework has examined the effects of different windowing
functions in detail \citep{Sundaresan2012, Whelan2015}.

Other methods can also be used to reduce spectral leakage. These include over-resolving each SFT by zero-padding (although this can still lead to some spectral leakage for signals in which frequency varies continuously with time), sinc-interpolating between SFT bins (thus leveraging the sampling theorem \cite{Shannon1998}), or simply adding contributions from neighboring SFT bins.  Including just the two adjacent SFT bins when cross-correlating can improve recovery of the expected SNR from $\approx$77\% to $\approx$90\% \cite{Sundaresan2012}.

In what follows, we acknowledge that spectral leakage could lead to SNRs that
are roughly $\approx$75\% of the idealized value for the $\rho$ statistic (i.e. up to a factor of $\sqrt{.75}\approx87\%$ in signal amplitude and/or distance reach for cases in which $\ddot{f}$ and higher terms may not be negligible). This is consistent with the estimate of $77.4\%$ for rectangular windowing described in \cite{Sundaresan2012} and related searches, e.g. \cite{Whelan2015}. Signals for which a chosen baseline is particularly close to the limit set by $\dot{f}$ via the quarter-cycle criterion (assumption \#2 in Sec. \ref{signalsft}) may experience additional leakage (not to exceed the maximum loss of $\sqrt{.5}\approx71\%$ in amplitude sensitivity, see Fig. \ref{fig:leakage-1-a}).


\section{GRB plateau search sensitivity}
\label{sec:plateau}
In this Section we apply the cross-correlation statistic to the specific model of intermediate-duration GW signals described in \cite{Corsi2009}. This model describes the scenario of a secularly unstable GRB-magnetar possibly associated with a GRB afterglow plateau (see also Section \ref{sec:intro}).

In the Newtonian limit, the $l=m=2$ $f$-mode becomes secularly unstable when the ratio $\beta=T/|W|$ of the rotational kinetic energy $T$ to the gravitational binding energy $|W|$ is between $0.14$ and $0.27$. This mode has the shortest growth time of all polar fluid modes, $1~{\rm s} \lesssim\tau_{GW}\lesssim 7\times10^4$ s for $0.24\gtrsim\beta\gtrsim0.15$ \citep{LaiShapiro1995} and may be an important source of GWs. Under the hypothesis that a secular bar-mode instability does indeed set in for a magnetar left over after a GRB explosion, \citet{Corsi2009} have  followed the NS quasi-static evolution under the effect of gravitational radiation according to the analytical formulation given by \citep{LaiShapiro1995}. Since $\tau_{GW}$ is generally much longer than the dynamical time of the star, the evolution is quasi-static, i.e., the star evolves along an equilibrium sequence of Riemann-S ellipsoids. Differently from what was done by \citet{LaiShapiro1995}, \citet{Corsi2009}  added into the evolution energy losses due to magnetic dipole radiation, assuming that those will not substantially modify the dynamics, but will act to speed up the overall evolution along the same sequence of Riemann-S ellipsoids that the NS would have followed in the absence of radiative losses.

\begin{figure}[htbp]
\begin{center}
\includegraphics[width=0.5\textwidth]{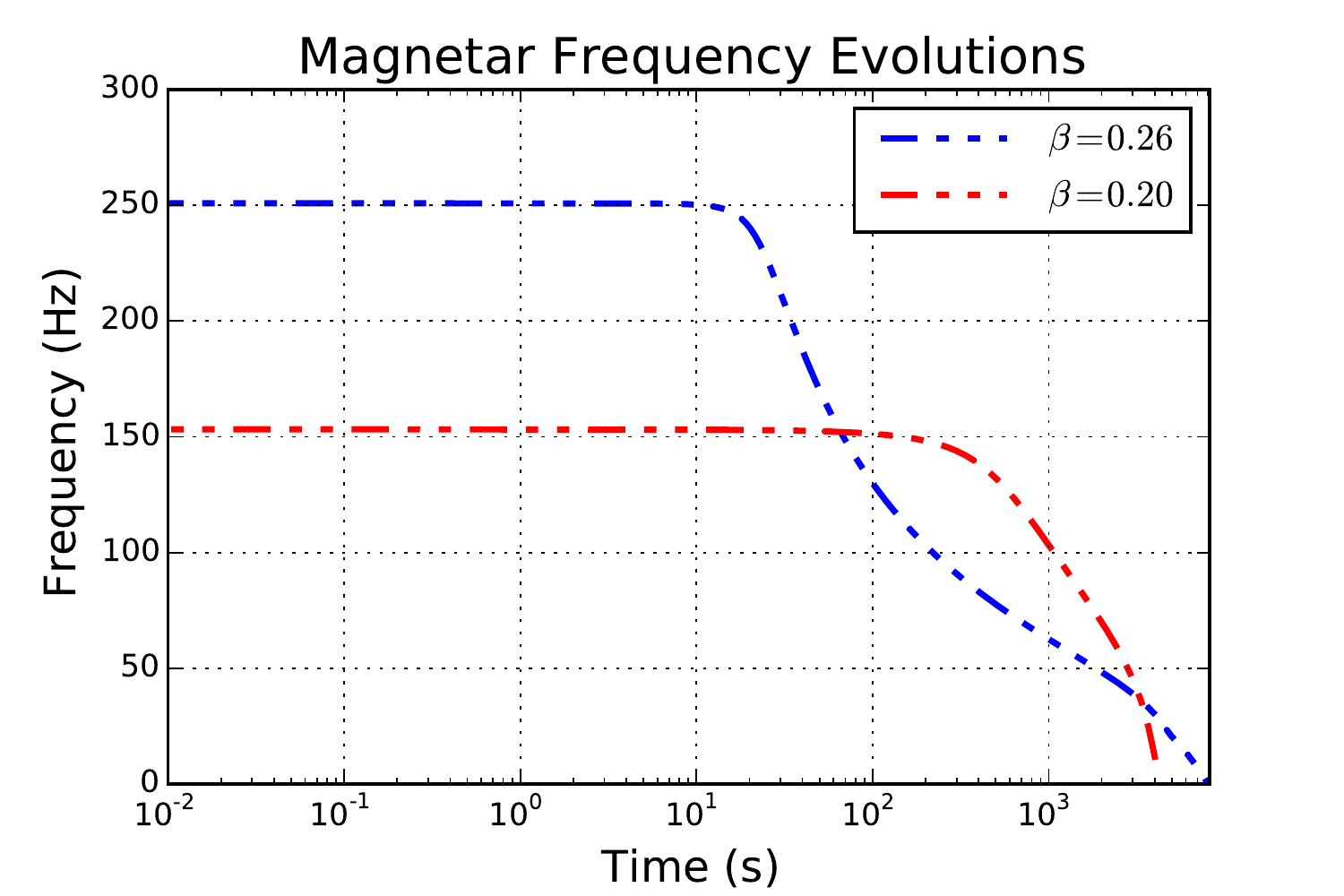}
\caption{Frequency evolution for two representative signals generated via the Corsi and M\'esz\'aros model. The signals are for a typical choice of parameters $M=1.4$ M$_{\odot}$, $R_0=20$\,km, $n=1$, $B_0=10^{14}$\,G, and different $\beta$ ($\beta=0.20$ is in the center of the allowed range of $0.14<\beta<0.27$). For more details on the model used to generate these waveforms, see \cite{Corsi2009}.}
\label{fig:freq-evolution}
\end{center}
\end{figure}

In the model proposed by Corsi \& M\'esz\'aros 2009 \cite{Corsi2009}, the resulting quasi-periodic GW signal depends on five parameters: $\beta$, the initial kinetic-to-gravitational potential energy ratio of the magnetized NS \cite{LaiShapiro1995}; $n$, the NS polytropic index; $M$, the NS mass; $R_0$, the unperturbed NS radius; and $B_0$, the  initial dipolar magnetic field strength at poles.  For a typical parameter choice of $M=1.4$ M$_{\odot}$, $R_0=20$\,km, $n=1$, $B_0=10^{14}$\,G, and $\beta=0.20$ (Fig. \ref{fig:freq-evolution}, red), \cite{Corsi2009} have estimated a distance reach (assuming a matched filter search) of $\approx$100\,Mpc for the aLIGO-Virgo detectors (for FAP$\approx5\times 10^{-5}$ and FDP=50\%).

We have tested the detectability of this class of signals using the adaptation of the cross-correlation statistic described in the previous Sections, and assuming Gaussian noise with $S_n(f)\approx1.83\times10^{-47}\,\rm{Hz}^{-1}$  approximately equal to that of whitened aLIGO-Virgo noise in the signal's frequency band (Fig. \ref{fig:noise-approximation}). The results are reported in Table \ref{tab:no-step} for an optimally oriented GRB\footnote{Here, ``optimally oriented'' is taken to mean that the GRB jet is aligned with the line of sight (so that $\iota=0$ and the GW is circularly polarized, i.e. $\A_+=\A_\times=1$, see Sec. \ref{signalsft}) and the GRB sky location is such that the line of sight is orthogonal to the plane containing the detector (so that  $F^2_++F^2_\times=1$).}. 

A matched filter analysis yields the highest sensitivity, and thus the largest horizon distance limits. For a typical choice of model parameters (e.g. $\beta=0.20$,  $B_0=10^{14}\,\rm{G}$, $M=1.4\,M_\odot$, $R_0=20\,\rm{km}$), if we assume that the initial phase is known as in \cite{Corsi2009}, we obtain a distance limit of $\approx$139\,Mpc for a FAP of 0.1\% and a FDP of 50\% using data from a single detector. This is consistent with the estimate of $\approx$100\,Mpc reported in \citep{Corsi2009} (which assumed a smaller FAP). A real matched filter search will have an overall unknown phase constant (see the last paragraph in Sec. \ref{Sec:matched}), which reduces the horizon distance to  $\approx$\mflim~ (for the same model parameters). Our cross-correlation matched filter limit yields an horizon distance of $\approx$103\,Mpc, or $103\,\rm{Mpc}/\sqrt{75\%}\approx$119\,Mpc when correcting for spectral leakage (see Fig. \ref{fig:leakage-1}), in agreement with the (real) matched filter.

We finally note that signals with faster frequency evolutions are affected more by spectral leakage, for a fixed choice of $\baseline$ (satisfying the quarter cycle criterion). For example, the GW signal from a magnetar with $\beta=0.26$ would have a faster frequency evolution than that from a source with $\beta=0.20$ (with other source parameters unchanged; Fig. 7). The distance horizon we achieve in the cross-correlation matched filter limit for $\beta=0.26$ (and for a $\baseline$ equal to the one used for the $\beta=0.20$ case) is $\approx 216$\,Mpc. This is $\approx\sqrt{57\%}$ of the expected matched filter horizon of 287 Mpc (see Table I), worse than what we would have expected for average spectral leakage losses of $\sqrt{75\%}$, but still within the maximum expected value of $\sqrt{50\%}$ (see Fig. 6a). Such extremal losses are consistent with baselines very near to (but not exceeding) the maximum value set by the quarter-cycle criterion. These losses can be improved by optimizing the size of the baseline, given each frequency evolution, and is planned for future work. These results are summarized in Table \ref{tab:no-step}.

\begin{table}[thbp]
\caption{Single-detector distance horizons for simulations in which the search is performed on the ``correct'' frequency-time track
for with $B_0=10^{14}\,\rm{G}$, $M=1.4\,M_\odot$, $R_0=20\,\rm{km}$ and varying values of $\beta$ using the model proposed by \cite{Corsi2009}. The search techniques used are 
matched filtering \emph{with unknown phase} (MF), the cross-correlation matched filter limit ($\chi^2$MF, see Sec. \ref{Sec:matched}), and the cross-correlation
stochastic limit (see Sec. \ref{Sec:stochastic}).}
\begin{center}
\begin{tabularx}{\columnwidth}{XXXX}
\hline\hline
\multirow{2}{*}{$\beta$ Value} 	&	\multicolumn{3}{c}{Distance Horizon (Mpc)}		\\
						&	MF		&	$\chi^2$MF	&	Stochastic		\\
\hline
0.20 						&	118		&	103			&	20			\\
0.26 						&	287		&	216			&	40			\\
\hline\hline
\end{tabularx}
\end{center}
\label{tab:no-step}
\end{table}%

While a detailed study of the parameter space of the model by \cite{Corsi2009} is beyond the scope of this paper, we also carried out several simulations to demonstrate the effectiveness of a semi-coherent approach in: (i) enhancing the robustness of the search against signal uncertainties when compared to a matched-filter limit; and (ii) enhancing the sensitivity of the search when compared to a ``stochastic approach''. We do so by calculating the distance horizons for situations in which the \emph{assumed} time-frequency track differs from the \emph{actual} signal by some amount. This difference is quantified by an error ($\delta M,~\delta R,~\delta B$) on the values of the true signal parameters ($M,~R,~B$). The sizes of these errors help determine the parameter space resolution for an effective search. The results of these tests are summarized in Tables \ref{tab:large-steps} and \ref{tab:small-steps}.

Because an error in signal parameters implies a mismatch between the true signal time-frequency evolution and the time-frequency track adopted for the calculation of the $\rho$\,statistic,  we expect the cross-correlation search to completely miss the signal in the limit of large coherence timescales, $\Tcoh\rightarrow\Tobs$ (approaching the matched-filter limit, which is not robust against such deviations). On the other hand, in the limit of small coherence timescales, $\Tcoh\rightarrow\baseline$, while the search is expected to be robust against signal uncertainties, the sensitivity is significantly lower than the matched-filter case. Thus, for a given parameter space resolution, one can define an optimal coherence timescale, which can then be used to quantify the distance reach of the semi-coherent regime (for given FAP and FDP).

We obtain the optimal coherence time ($T_\text{opt}$) by calculating the detection efficiency for given FAP (here, 0.1\%) as a function of $\Tcoh$, for a signal at a fixed distance. The $\Tcoh$ that is associated with the maximal detection efficiency is then used for a series of injections of varying distance, but fixed $\Tcoh$. The distance that is associated with an efficiency of 50\% (which is equivalent to a FDP of 50\%) is then taken to be the distance horizon for that step size. The step sizes taken for each model parameter informs the size of the parameter space that a semi-coherent cross-correlation search should cover. We ran simulations with two classes of step size: ``large'' steps that correspond to a coarse grid in the parameter space, and ``small'' steps that correspond to finer (and subsequently, more computationally intensive) grid in the parameter space.

\begin{figure*}[!htb]
\begin{center}
\includegraphics[width=\textwidth,angle=0]{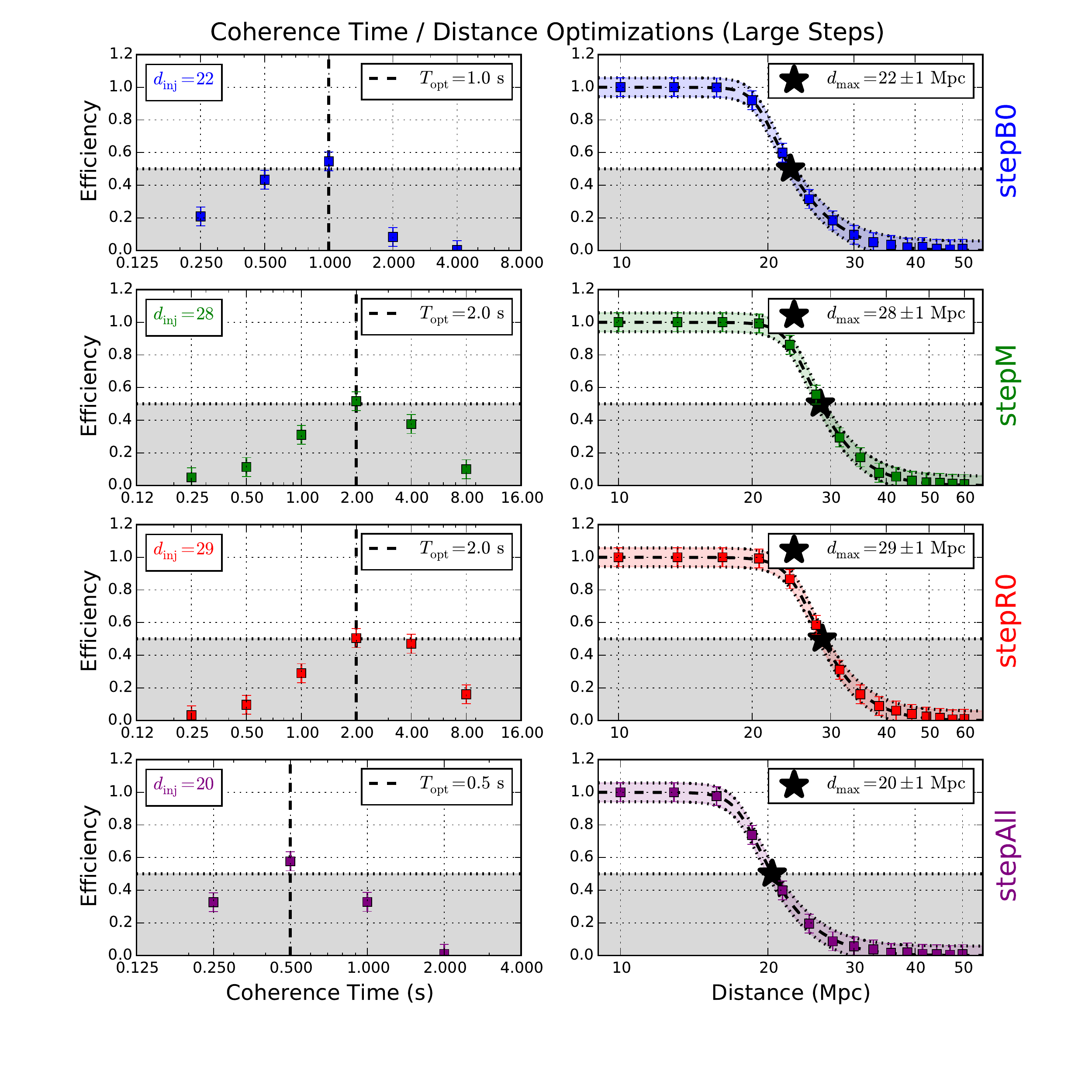}
\caption{Efficiency (1-FDP) plots for \textit{large} steps $\delta B_0=10^{12}\,\rm{G}$ (blue), $\delta M=5\times10^{-3}\,M_\odot$ (green), $\delta R_0=20\,\rm{m}$ (red) and all three combined (purple). All plots assume FAP=0.1\% and distances are extracted using FDP=50\% (black dotted line and gray shaded area). On the left, optimal coherence time plots. The signal is injected at a constant distance, $\Tcoh$ is then varied to find the value that maximizes detection efficiency ($T_{\rm{opt}}$). On the right, $\Tcoh$ is fixed at the optimum value for each step, and then distance is varied. The result is fit by an asymmetric sigmoid of the form $\rm{sig}(x)=[1+\exp(p_0\{x-p_1\})]^{-1/p_2}$ (where $p_0,~p_1,~p_2$ are constants to be fit), which is then used to interpolate and determine the max distance ($d_{\rm{max}}$).}
\label{fig:large_steps}
\end{center}
\end{figure*}

The results for the large steps are shown in Fig. \ref{fig:large_steps}, and summarized in Table \ref{tab:large-steps}. Optimal coherence times, see Fig. \ref{fig:large_steps} (left), are of $\mathcal{O}(1)\,\rm{s}$, which lead to maximal detection distances around $20-30$\,Mpc (recovering only $\approx 25$\% of the matched filter limit), see Fig. \ref{fig:large_steps} (right). In the case where all three parameters are stepped simultaneously ($\delta\text{All}$), the optimal coherence time is only twice the SFT baseline of $\baseline=0.25\,\rm{s}$ and provides no significant gains over the stochastic limit, see Table \ref{tab:large-steps}.

\begin{table}[thbp]
\caption{Single-detector distance horizons for \emph{large} steps in each of the model parameters with, $\beta=0.20$, $B_0=10^{14}\,\rm{G}+\delta B_0$, $M=1.4\,M_\odot+\delta M$, $R_0=20\,\rm{km}+\delta R_0$. The resulting distance horizons are approximately 20-30 Mpc, which is up to a 50\% improvement over the stochastic limit,\footnote{A factor of 1.5 in distance horizon increases the expected detection rate by a factor of $1.5^3\approx 3$.} but only $\approx$25\% of the matched filter limit. All errors of order $\mathcal{O}(1)$\,Mpc.}
\begin{center}
\begin{tabularx}{\columnwidth}{lXXXX}
\hline\hline
\multirow{2}{*}{Parameter}		& \multirow{2}{*}{Step size}&\multirow{2}{*}{$T_\text{opt}$ (sec)}	&	\multicolumn{2}{c}{Distance Horizon (Mpc)}\\
						&							&		&	Semicoh		&		Stochastic		\\
\hline
$\delta B_0$	&	$10^{12}\,\rm{G}$			&	1	&	22			&		20			\\
$\delta M$		&	$5\times10^{-3}\,M_\odot$		&	2	&	28			&		20			\\
$\delta R_0$	&	$20\,\rm{m}$				&	2	&	29			&		20			\\
$\delta\text{All}$&	As above					&	0.5	&	20			&		20			\\
\hline\hline
\end{tabularx}
\end{center}
\label{tab:large-steps}
\end{table}%

\begin{figure*}[!htb]
\begin{center}
\includegraphics[width=\textwidth,angle=0]{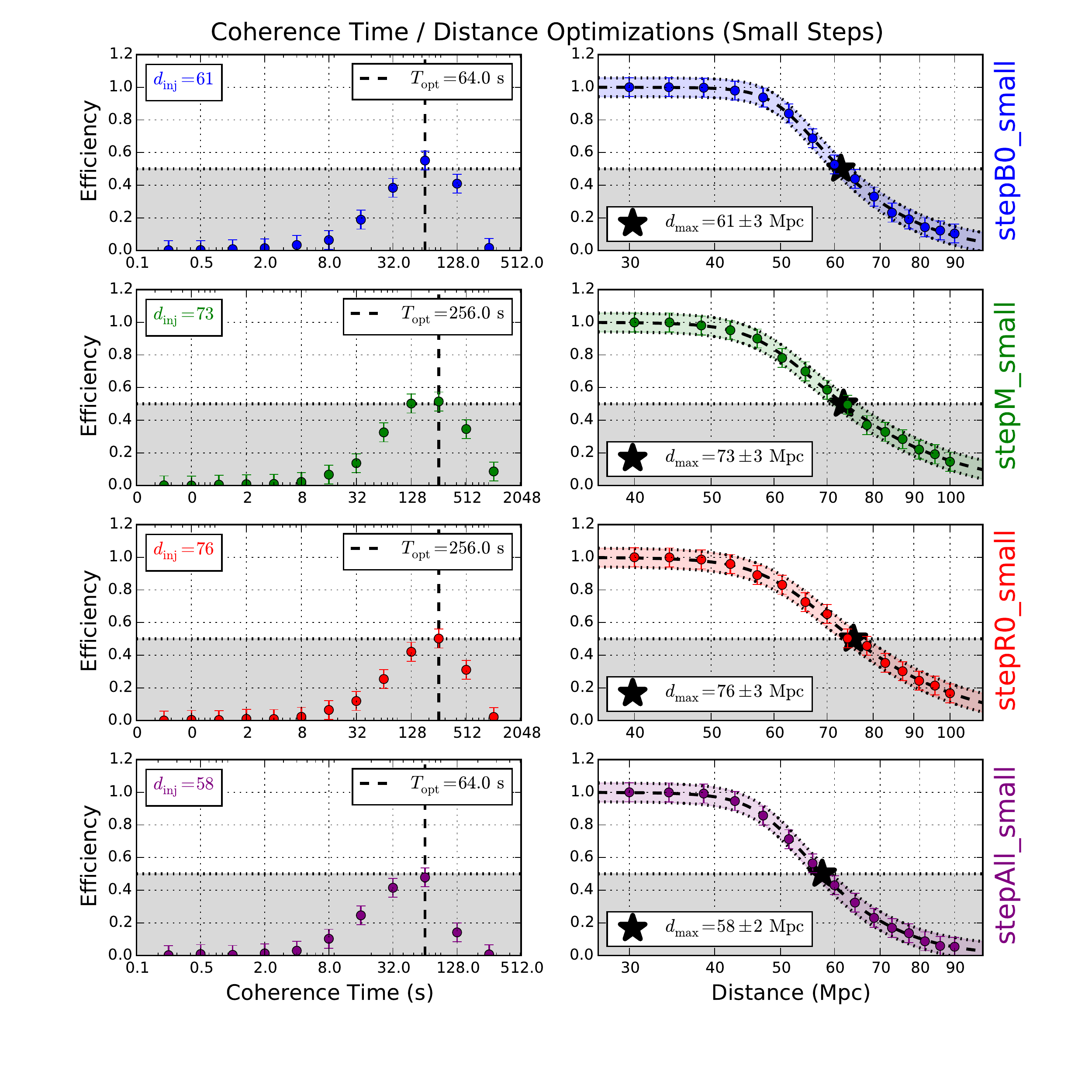}
\caption{Efficiency (1-FDP) plots for \textit{small} steps in $\delta B_0=10^{10}\,\rm{G}$ (blue), $\delta M=5\times10^{-5}\,M_\odot$ (green), $\delta R_0=0.2\,\rm{m}$ (red) and all three combined (purple). All plots assume FAP=0.1\% and distances are extracted using FDP=50\% (black dotted line and gray shaded area). On the left, optimal coherence time plots. The signal is injected at a constant distance, $\Tcoh$ is then varied to find the value that maximizes detection efficiency ($T_{\rm{opt}}$). On the right, $\Tcoh$ is fixed at the optimum value for each step, and then distance is varied. The result is fit by an asymmetric sigmoid of the form $\rm{sig}(x)=[1+\exp(p_0\{x-p_1\})]^{-1/p_2}$ (where $p_0,~p_1,~p_2$ are constants to be fit), which is then used to interpolate and determine the max distance ($d_{\rm{max}}$).}
\label{fig:small_steps}
\end{center}
\end{figure*}

The small step sizes produce optimal coherence times of as high as $256\,\rm{s}$, see Fig. \ref{fig:small_steps} (left), which lead to maximal detection distances of $\approx$60-80\,Mpc, roughly $\approx$75\% of the matched filter limit, see Fig. \ref{fig:small_steps} (right). For comparison, we note that nearest long GRB on record was GRB\,980425, located at a distance of $40$\,Mpc \cite{Tinney1998, Galama1998}. These results suggest that the large steps considered above are indeed too large to adequately resolve the parameter space, while the small steps represent a good starting point for a finer exploration of the physically relevant parameter space. We note that an in depth discussion of parameter space range and resolution must also include the effect of the implied number of trials on the detection statistic. This effect is expected to be more important for longer coherence times. A full study of the parameter space for intermediate-duration GWs, using the cross-correlation search technique described here, is planned for future work.

\begin{table}[thbp]
\caption{Single-detector distance horizons for \emph{small} steps in each of the model parameters with, $\beta=0.20$, $B_0=10^{14}\,\rm{G}+\delta B_0$, $M=1.4\,M_\odot+\delta M$, $R_0=20\,\rm{km}+\delta R_0$. The simulation used $\Tobs=1024$\,s and $\baseline=0.25$\,s. The resulting distance horizons are approximately 60-80 Mpc, up to four times as large as the stochastic limit,\footnote{A factor of 4 in distance horizon increases the expected detection rate by a factor of $4^3=64$.} and $\gtrsim$75\% of the matched filter limit. All errors of order $\mathcal{O}(1)$\,Mpc.}
\begin{center}
\begin{tabularx}{\columnwidth}{lXXXX}
\hline\hline
\multirow{2}{*}{Parameter}		& \multirow{2}{*}{Step size}&\multirow{2}{*}{$T_\text{opt}$ (sec)}	&	\multicolumn{2}{c}{Distance Horizon (Mpc)}\\
						&							&		&	Semicoh		&		Stochastic		\\
\hline
$\delta B_0$	&	$10^{10}\,\rm{G}$			&	64	&	61			&		20			\\
$\delta M$		&	$5\times10^{-5}\,M_\odot$		&	256	&	73			&		20			\\
$\delta R_0$	&	$0.2\,\rm{m}$				&	256	&	76			&		20			\\
$\delta\text{All}$&	As above					&	64	&	58			&		20			\\
\hline\hline
\end{tabularx}
\end{center}
\label{tab:small-steps}
\end{table}%


\section{Discussion and Conclusion}
\label{sec:discussion}

We have explored the application of the cross-correlation technique described in \citep{Dhurandhar2008} to a new class of intermediate
duration GW signals of duration $\Tobs\lesssim 10^4$\,s, specifically the bar mode instability model for millisecond
magnetars developed in \cite{Corsi2009}. In doing so, we have corrected the statistical properties of the cross-correlation
statistic reported in \cite{Dhurandhar2008} for both the semi-coherent, and fully-coherent matched-filter limits. In addition, we have done a cursory exploration of the parameter
space for this model.

There are several parallels between limits of the cross-correlation method and other search techniques used for LIGO data analysis. Natural examples are the techniques derived from efforts to quantify the stochastic GW background. Two such methods are the Stochastic Transient Analysis Multi-detector Pipeline (STAMP), a cross-power statistic widely used for LIGO all-sky searches \cite{Thrane2011, Coughlin2011}, and stochtrack, a seedless clustering algorithm that has been tested on signal models comparable in duration to those considered here \cite{Thrane2014}. Both these methods are similar (in spirit, if not necessarily implementation) to the stochastic limit of the cross-correlation approach.

Because of their significant robustness against signal uncertainties (and relatively low computational costs) stochastic-inspired methods (as the two described above) are attractive for many search regimes, and especially as a first pass when searching for viable GW candidates with wide parameter spaces. On the other hand, the improvement in sensitivity (and therefore distance reach) enabled by the semi-coherent limit of the cross-correlation approach lends itself to deeper searches. A potential way to leverage the strengths of both regimes is to develop a framework in which a stochastic-inspired search is used for discovery, with semi-coherent cross-correlation followup for parameter estimation and refinement. This could be done entirely within cross-correlation method described in this work, or by using an established stochastic technique (e.g. STAMP) for discovery and cross-correlation for follow-up.

Overall, the results of our study are encouraging: The tunable robustness versus sensitivity of the cross-correlation technique is well suited for intermediate-duration GW signals that evolve on timescales of $10^3$-$10^4$\,s, and can reach astrophysically relevant distance horizons with the expected noise characteristics of  GW detectors such as aLIGO and Virgo. However, a full parameter space exploration is yet to be completed, as is testing on real instrument noise. Additionally, the trials factor for a full parameter space search will reduce, to some extent, the idealized horizon distances calculated here. We intend to explore these aspects of the analyses in future work.

\acknowledgments
This work is supported by NSF grant PHY-1456447 (PI: Corsi). B.O. acknowledges
support from NSF grants PHY-1206027, PHY-1544295, and PHY-1506311. A.C. and
B.O. thank P. Meszaros for useful discussions in the early stages of this
work. A.C. also thanks C. Palomba for early discussions regarding continuous wave searches. This paper has been assigned LIGO Document No,
LIGO-P1500226.

\bibliographystyle{apsrev-titles}
\bibliography{coyne_corsi_owen_2015}

\end{document}